

\input amstex
\input amsppt.sty            
\magnification=1000		
\pagewidth{6.  truein}
\pageheight{9. truein}
\hoffset = .15in
\voffset = 0in
\NoBlackBoxes
\TagsOnRight

\topmatter
\headline={\hfill \the \pageno}
\title
Kinematic Self-Similarity
\endtitle

\author
A. A. Coley \endauthor
\affil  Department of Mathematics\\
Statistics and Computing Science\\
Dalhousie University\\
Halifax, Nova Scotia\\
Canada B3H 3J5
\endaffil

\abstract
Self-similarity in general relativity is briefly reviewed and the differences 
between self-similarity of the first kind (which can be obtained from dimensional
 considerations and is invariantly characterized by the existence of a homothetic
 vector in perfect fluid spacetimes) and generalized self-similarity are 
discussed.  The covariant notion of a kinematic self-similarity in the context of
 relativistic fluid mechanics is defined.  It has been argued that kinematic 
self-similarity is an appropriate generalization of homothety and is the natural 
relativistic counterpart of self-similarity of the more general second (and 
zeroth) kind.  Various mathematical and physical properties of spacetimes 
admitting a kinematic self-similarity are discussed.  The governing
equations for perfect fluid cosmological models are introduced and 
a set of integrability conditions for the existence of a proper
kinematic self-similarity in these models is derived.  Exact
solutions of the irrotational perfect fluid Einstein field equations
admitting a kinematic self-similarity are then sought in a 
number of special cases, and it is found that; (1) in the geodesic 
case the $3$-spaces orthogonal to the fluid velocity vector are necessarily
Ricci-flat and (ii) in the further specialisation to dust (i.e., zero pressure)
the differential equation governing the expansion can be completely
integrated and the asymptotic properties of these solutions can be 
determined, (iii) the solutions in the case of zero-expansion consist
of a class of shear-free and static models and a class of stiff perfect
fluid (and non-static) models, and (iv) solutions in which the kinematic
self-similar vector is parallel to the fluid velocity vector are necessarily
Friedmann-Robertson-Walker (FRW) models.  Solutions in which the
kinematic self-similarity is orthogonal to the velocity vector are also
considered.  In addition, the existence of kinematic self-similarities
in FRW spacetimes is comprehensively studied.  It is known that 
there are a variety of circumstances in general relativity in which self-similar
models act as asymptotic states of more general models.  Finally, the 
questions of under what conditions are models which admit a proper
kinematic self-similarity asymptotic to an exact homothetic solution 
and under what conditions are the asymptotic states of 
cosmological models represented by exact solutions of Einstein's 
field equations which admit a generalized self-similarity are addressed. 
\endabstract

\endtopmatter

\document
\baselineskip11pt  

\heading{1. Introduction} \endheading

Self-similar solutions were originally of 
interest since the governing equations of a given problem
simplify and often systems of partial differential
equations (PDEs) reduce to ordinary differential equations.
Indeed, self-similarity in the broadest (Lie) sense refers to 
an invariance which simply allows the reduction of a system of PDEs. 
Self-similarity refers to the fact that the spatial 
distribution of the characteristics of motion remains
similar to itself at all times during the motion and self-similar solutions represent solutions of degenerate problems 
in which all dimensional constant parameters entering
the initial and boundary conditions vanish or become
infinite (Barenblatt and Zeldovich, 1972). Indeed, such solutions 
describe the ``intermediate-asymptotic'' behaviour of 
solutions in the region in which a solution
no longer depends on the details of the initial and/or
boundary conditions.  Cases in which the form of the self-similar asymptotes 
can be obtained from dimensional considerations are referred to as 
\underbar{self-similar solutions of the first kind} (Barenblatt and Zeldovich, 1972).

Similarity solutions are  of importance within general relativity.  
For example, a strong explosion in a homogeneous background produces
fluctuations which may be very complicated initially, but they
tend to be described more and more closely by a spherically
symmetric similarity solution as time evolves (Sedov, 1967), and this
applies even if the explosion occurs in an expanding cosmological 
background (Schwartz et al., 1975; Ikeuchi et al., 1983).  The evolution of voids is also described by similarity solutions at
late times (Bertschinger, 1985).  In addition, 
the expansion of the universe from the big
bang and the collapse of a star to a singularity might (both)
exhibit self-similarity in some form since it might be expected
that the initial conditions `are forgotten' in some sense.      

In this paper we shall assume that the
source of the gravitational field is that of a perfect fluid; i.e.,
the energy-momentum tensor is given by
$$T_{ab} = (\mu + p) u_a u_b + pg_{ab}, \tag1.1$$
where $u^a$ is the normalized fluid $4$-velocity and $\mu$ and $p$ are,
respectively, the density and pressure.  In natural units $c = 8 \pi G = 1$,
the Einstein field equations (EFEs) of general relativity then read
$$G_{ab} = T_{ab}, \tag1.2$$
where $G_{ab}$ is the Einstein tensor.

Similarity solutions were first studied
in general relativity by Cahill and Taub (1971), who  did so in the cosmological context and 
under the assumption of a spherically symmetric distribution of a self-gravitating perfect fluid. They assumed that the solution was
such that the dependent variables are essentially functions of a single
independent variable constructed as a \underbar{dimensionless} combination of the
independent variables and that
the model contains no
other dimensional constants.
Cahill and Taub (1971) showed that the existence of 
a similarity  of the first kind in this situation could be 
invariantly 
formulated in terms of the existence of a homothetic vector.  A proper homothetic
vector (HV) is a vector field $\pmb{\xi}$ which satisfies (after a constant rescaling)
$${\Cal L}_{\pmb{\xi}} g_{ab} = 2g_{ab}, \tag1.3$$
where $g_{ab}$ is the metric and ${\Cal L}$ denotes Lie differentiation along $\pmb{\xi}$.  
 
It follows from equation (1.3) that
$$  {\Cal L}_{\pmb{\xi}} G_{ab} = 0. \tag1.4 $$
When the source of the gravitational field is a
perfect fluid, in the case of self-similarity
of the first kind it follows from dimensional considerations that the physical quantities transform according
to 
$${\Cal L}_{\pmb{\xi}} u^a = -u^a, \tag1.5 $$
and
$${\Cal L}_{\pmb{\xi}} \mu = - 2 \mu, \enskip {\Cal L}_\xi p = -2p. \tag1.6 $$
{}From these equations it follows that
$${\Cal L}_{\pmb{\xi}} T_{ab} = 0, \tag1.7 $$
which is therefore consistent with the EFEs (1.2).
Indeed, in the case of a perfect fluid it follows that
equations (1.5) and (1.6) result from equations (1.3) [through eqns. (1.2), (1.4) and (1.7)] so that the physical
quantities transform appropriately (Cahill and Taub, 1971; Eardley, 1974).  Hence in this case ``geometric'' self-similarity
and ``physical'' self-similarity coincide.  However, this need not be the case (see Coley, 1996, for details).  The
properties of the matter and those of the geometry are related 
through the EFEs, and in general there will be further constraints arising from the compatibility
of the EFEs and the conditions of self-similarity (``integrability'' conditions).
We shall investigate these integrability conditions later.

\subhead{A.  Self-similarity of the second kind}\endsubhead 

The existence of self-similar solutions of the first type
is related to the conservation laws and to the invariance of 
the problem with respect to the group of similarity transformations
of quantities with independent dimensions, in which case
a certain regularity of the limiting
process in passing from the original non-self-similar regime to the
self-similar regime is implicitly assumed.  However, in general
such a passage to this limit need not be regular, whence the
expressions for the self-similar variables are not determined 
from dimensional analysis of the problem alone.  Solutions
are then called {\it self-similar solutions of the second type}.
Characteristic of these solutions is that they {\it contain dimensional constants} that are not determined from the conservation 
laws (but can be found by matching the self-similar
solutions with the non-self-similar solutions whose
asymptotes they represent) (Barenblatt and Zeldovich, 1972).

Self-similarity in the broadest sense refers to the general situation in 
which a system is not restricted to be strictly invariant under the relevant group action, but
merely to be appropriately rescaled.  The basic condition characterizing a manifold vector field
$\pmb{\xi}$ as a self-similar generator (Carter and Henriksen,1991) is that there exist constants $d_i$
such that for each independent physical field $\Phi^i_A$,
$${\Cal L}_{\pmb{\xi}} \Phi^i_A = d_i \Phi^i_A ,\tag1.8$$
where the fields $\Phi^i_A$ can be scalar (e.g., $\mu$), vectorial (e.g., $u_a$) or
tensorial (e.g., $g_{ab}$).  In general relativity
the gravitational field is represented by the metric
tensor $g_{ab}$, and an appropriate definition of ``geometrical''
self-similarity is necessary.  In the seminal work by
Cahill and Taub (1971) the simplest generalization was 
effected whereby the metric itself satisfies an equation
of the form (1.8), namely $\pmb{\xi}$ is a HV, this evidently
corresponding to Zeldovich's similarity of the first kind.

However, in relativity it is not the energy-momentum tensor itself 
that must satisfy (1.8), but each of the 
physical fields making up the energy-momentum tensor
must separately satisfy an equation of the form (1.8).
In the case of a fluid characterized by the 
timelike congruence $u_a$ the energy-momentum tensor can be
uniquely decomposed with respect to $u_a$ (Ellis, 1971),
and each of these uniquely defined components
(each of which has a physical interpretation in terms of
the energy, pressure, heat flow and anisotropic stress as
measured by an observer comoving with the fluid)
must separately satisfy an equation of the form (1.8).
In the same way, if the metric can be uniquely,
physically, and covariantly decomposed then the homothetic condition can
be replaced by the conditions that each uniquely defined
component must satisfy (1.8), maintaining self-similarity.
For example, in the case of a fluid, the metric can be decomposed
uniquely in terms of $u_a$, through the projection tensor
$$h_{ab} = g_{ab} + u_a u_b, \tag1.9 $$
into parts $h_{ab}$ and (minus) $u_a u_b$.  The projection tensor
represents the projection of the metric into
the $3$-spaces orthogonal to $u^a$ (i.e., into the 
rest frame of the comoving observers), and if $u_a$ is irrotational
these $3$-spaces are surface forming, the decomposition
is global, and $h_{ab}$ represents the intrinsic metric of
these $3$-spaces.  [$h_{ab}$ is the first fundamental form of the hypersurfaces orthogonal to $u^a$ and 
can be regarded as the
relativistic counterpart of the Newtonian metric tensor,
when the flow independent $u_a$ is defined as the relativistic counterpart 
of the preferred (\underbar{irrotational})
Newtonian time covector $-t_{,a}$ (Carter and Henriksen, 1989).]

\subhead{B. Kinematic self-similarity}\endsubhead

It is arguments similar to these, and, more
importantly, a detailed comparison with 
self-similarity in a continuous Newtonian medium, that has led Carter and Henriksen (1989)
to the covariant notion of {\it kinematic self-similarity} in the
context of relativistic fluid mechanics.  A kinematic
self-similarity vector $\pmb{\xi}$ satisfies the condition
$${\Cal L}_{\pmb{\xi}} u_a = \alpha u_a, \tag1.10$$
where $\alpha$ is a constant (i.e., $\pmb{\xi}$ is a continuous kinematic self-similar
generator with respect to the flow $u_a$). Furthermore,
$${\Cal L}_{\pmb{\xi}} h_{ab} = 2h_{ab} \tag1.11$$
($\pmb{\xi}$ has been normalized so that the constant in (1.11)
has been set to unity).  Evidently, in the case $\alpha =1$ 
it follows that $\pmb{\xi} $ is a HV (Cahill and Taub, 1971), corresponding
to self-similarity of the first kind (Barrenblatt and Zeldovich, 1972).
Carter and Henriksen (1989) then argue that the case $\alpha \neq 1$
$(\alpha =0)$ is the natural relativistic counterpart of self-similarity
of the more general second kind (zeroth kind).

The parameter $\alpha$ represents the constant relative proportionality
factor governing the rates of dilation of spatial length scale and amplification
of time scale.  Evidently, when $\alpha \neq 1$ (i.e., $\pmb{\xi}$ is not a HV), the
relative rescaling of space and time (under ${\pmb \xi}$) are not the same
(and in the zeroth case there is a space
dilation without any time amplification).  

\subhead{C. Remarks}\endsubhead

1. All of the fields must satisfy equations of the form (1.8).
However, not all of the $d_i$ need be independent.  In practice
it is convenient to assume independence and then
determine the constraints on the $d_i$ (in addition to the constraints on the form of the solutions)
that arise from the imposition of the equations of state and the
EFEs.  In the case of a perfect fluid we have that, in
addition to the conditions of ``kinematic'' self-similarity and 
``geometric'' self-similarity, as represented by equations (1.10) and (1.11), respectively, the further conditions of ``physical'' self-similarity, given by 
$$ {\Cal L}_{\pmb{\xi}} \mu = a \mu, \enskip {\Cal L}_{\pmb{\xi}} p = bp , \tag1.12 $$
must be satisfied,
where $a$ and $b$ are constants.  That is, in order for a perfect fluid spacetime to admit a proper kinematic self-similarity all of the 
conditions represented by 
equations (1.10)--(1.12) must be satisfied.

2. Additional constraints may arise from the imposition of an equation of state.
The least restrictive case arises from the exceptional
pressure-free case.  Here $p=0$, whence 
$${\Cal L}_{\pmb{\xi}} T_{ab} = (2\alpha +a) T_{ab}, \tag1.13 $$
and no further restrictions apply.  
Usually  a linear barotropic equation of 
state of the form
$$p = (\gamma -1) \mu, \tag1.14 $$
is assumed, where the constant $\gamma$ obeys $1 \leq \gamma \leq 2$ for
ordinary
matter.  [The case $0 \leq \gamma < 2/3$ is of interest in studying models
that undergo inflation.]
If a barotropic equation of state $p = p(\mu)$ is assumed,
it is well known that if a spacetime admits a non-trivial HV
then equation (1.14) necessarily results (Cahill and Taub, 1971).
If, on the other hand, a spacetime admits a non-trivial 
kinematic self-similarity then (except in the special case of dust)
the polytropic equation of state $p = p_o \mu^{\overline{\gamma}}$ follows from equations (1.12) (where $\overline{\gamma}$ is the polytropic index
and ${\overline{\gamma}}$ = b/a).

We note from equation (1.13) that in the case of dust the total energy-momentum 
tensor $T_{ab}$ itself trivially satisfies an equation of the form (1.8).  When $p \neq 0$, this is only be possible in the special case that $b = a + 2 (\alpha -1)$.

3. The differential geometric properties of HV were studied by Yano (1955).  The
totality of HV on a spacetime form a Lie algebra $H_n$ (of dimension $n$)
which contains
(if $H_n$ is non-trivial) an $(n-1)$ dimensional
(sub)algebra of KV,
$G_{n-1}$.  If a given
spacetime is not conformal to an exceptional `plane wave spacetime', it follows
that if the orbits of $H_n$ are $r$-dimensional, then the
orbits of $G_{n-1}$ are
$(r-1)$-dimensional (Eardley, 1974), and if a spacetime is not conformally
flat,
the spacetime is conformally related to a spacetime for which the
Lie algebra $H_n$ is the Lie algebra of KV (Defrise-Carter, 1975).

The totality of kinematic self-similar vector fields 
on a spacetime also form a Lie algebra (Carter and 
Henriksen, 1991) which we shall denote here by {\it K}.  Now,
$H \subseteq K$, but $K$ neither contains nor is contained within
either the conformal algebra $C$ in general or the 
inheriting algebra $I$ (Castejon-Amenedo and Coley, 1992) in particular.

Let us suppose that $\pmb{\xi}_1$ and $\pmb{\xi}_2 \in K$, so that
$$\leqalignno{& {\Cal L}_{\pmb{\xi}_1} h_{ab} = 2h_{ab}, \qquad {\Cal L}_{\pmb{\xi}_1} u_a = \alpha_1 u_a,\cr
& {\Cal L}_{\pmb{\xi}_2}h_{ab} = 2h_{ab}, \qquad {\Cal L}_{\pmb{\xi}_2} u_a = \alpha_2 u_a,\cr}   $$
where $\pmb{\xi}_1$ and $\pmb{\xi}_2$ are linearly independent and $\alpha_1 \neq 1$ and $\alpha_2 \neq 1$ (i.e., neither are HV).  Then, by defining
$$\pmb{\xi}= c \pmb{\xi}_1 + d \pmb{\xi}_2, \tag1.15$$
where $c$ and $d$ are non-zero constants, we find that
$${\Cal L}_{\pmb{\xi}} h_{ab} = 2(c+d) h_{ab} \tag 1.16$$
and
$${\Cal L}_{\pmb{\xi}} u_a = (c \alpha_1 + d \alpha_2) u_a. \tag1.17 $$
[We note that equations of the form (1.12) are trivially satisfied for $\pmb{\xi}$ given by (1.15)].  Now, if $\alpha_1 \neq \alpha_2$ (where either $\alpha_1$ or $\alpha_2$ may be zero), then we can always choose $d=1 -c$ and $c = (1 - \alpha_2)/(\alpha_1 - \alpha_2)$ so that
${\Cal L}_{\pmb{\xi}} h_{ab} = 2h_{ab}$ and ${\Cal L}_{\pmb{\xi}} u_a = u_a$; that is, $\pmb{\xi}$ 
is a HV.  If, on the other
hand, $\alpha_1 = \alpha_2$ (possibly zero), then we can choose $c +d =0$,
whence ${\Cal L}_{\pmb{\xi}} h_{ab} =0$ and ${\Cal L}_{\pmb{\xi}} u_a =0$ so that $\pmb{\xi}$ is, in fact, a KV.
Therefore, we have shown that each non-trivial $K_n (n >1)$ contains 
an $(n-1)$-dimensional 
(sub)algebra of HV, $H_{n-1}$, where $H_{n-1}$ 
may be trivial (i.e., $H_{n-1}$ need not contain any \underbar{proper} HV).  An illustration of this result can be found in section 4.C.

4. In the case of radiation (i.e., $p = \frac{1}{3} \mu$, so
that $T= 0$)
the existence of a HV implies the existence of a conserved quantity.
In general, if we define the current  $P^a = T^{ab} \xi_b$, then from
energy-momentum
conversation, $P^a \,_{;a} = T^{ab} g_{ab} \equiv T$.  For radiation,
$P^a = \frac{\mu}{3}( 4 u^a [u_b \xi^b] + \xi^a)$, and $P^a \,_{;a} = 0$.
In the case of kinematic self-similarity it follows from equations (1.10)
and (1.11) that $P^a\,_{;a} = 3 p - {\alpha} {\mu}$, implying the existence
of a conserved quantity for fluids with an equation of state $p = \frac{\alpha}{3} \mu$. In particular, in the case of kinematic self-similarity of the zeroth
kind (i.e., $\alpha = 0$), there exists a conserved quantity in the special
case of dust (i.e., $p = 0$).

5. Finally, there is another case of potential interest in 
which the ratio $\alpha/1$ (where $\alpha$ is defined in (1.10) and we
recall that the constant in (1.11) has been normalized to
unity) approaches infinity, and we could refer to this
case as kinematic self-similarity of `infinite' kind.  This case 
could be covariantly defined by equation (1.10)
(in which $\alpha$ could be normalized to unity) and equation (1.11)
with zero right-hand side (i.e., ${\Cal L}_{\pmb{\xi}} h_{ab} = 0$), and $\pmb{\xi}$
consequently represents a generalized ``rigid motion''.  This
case will be investigated elsewhere.

In section 2 we shall describe the cosmological models 
under investigation and introduce their governing equations.  We shall 
then obtain the set
of integrability conditions for the existence of a kinematic  self-similarity in the models.  Finally, we shall deduce the equations in the relevant 
case of zero vorticity.  In sections 3 and 4 we shall study the models 
in a number of special cases; namely, in the case of zero 
acceleration (3.A) and the further subcase of zero pressure (3.B), 
the case of zero expansion (which includes the special subcase of 
static models) (3.C),  the cases in which   the kinematic self-similarity
is either parallel to or orthogonal to the fluid velocity vector (4.A 
and 4.B, respectively) and finally we shall investigate 
the existence of kinematic self-similar vectors in FRW spacetimes.
In section 5 we shall summarize the main results and we shall 
discuss further avenues of research.

\heading{2.  Analysis}\endheading

\subhead{A.  The governing equations}\endsubhead

The covariant derivative of $u_a$ can be decomposed 
according to (Ellis, 1971)\footnote"$^\dag$"{We shall follow the notation and conventions in Ellis (1971); in 
particular, Roman indices range from 0 to 3 and Greek indices from 1 to 3.}
$$u_{a;b} = \sigma_{ab} + \frac{1}{3} \theta h_{ab} + \omega_{ab} - \dot{u}_a u_b, \tag2.1 $$
where $\theta_{ab} \equiv h_{(a} \, ^c\!h_{b)} \,\!\!^d \, u_{c;d}, \theta \equiv g^{ab} \theta_{ab}$ is the expansion, 
$\sigma_{ab} = \theta_{ab} - \frac{1}{3} \theta h_{ab}$ is the shear tensor and $\sigma^2 \equiv \frac{1}{2} \sigma_{ab} \sigma^{ab}$, $\omega_{ab} \equiv h_{[a} \,^c\!h_{b]} \,\!\!^d u_{c;d}$ is the vorticity tensor and $\omega^2 \equiv \frac{1}{2} \omega_{ab} \omega^{ab}$, 
and $\dot{u}_a \equiv u_{a;b} u^b$ is the acceleration.$^\dag$  Using these definitions the governing equations can be written down (Ellis, 1971).

The conservation laws, in the case of a perfect fluid, become
$$\align
& \dot{\mu} + (\mu + p) \theta =0, \tag2.2\\
& (\mu +p) \dot{u}_a = -p_{, b}h^b\,_a,\tag2.3
\endalign $$
where $\dot{\mu} \equiv \mu_{,a}u^a$ (for example).
Equations (4.12), (4.15), (4.17), (4.18), (4.16) and (4.19) in
Ellis (1971), become, respectively,
$$\align
&\dot{\theta} + \frac{1}{3}\theta^2 - \dot{u}^a \,_{;a} + 2 (\sigma^2 - \omega^2) + \frac{1}{2} (\mu + 3p) = 0, \tag2.4\\
&  h^a\,_b\left( \frac{(l^2 \omega^b)^o}{l^2}\right)   = h^a\,_b\left(\dot{\omega}^b + \frac{2}{3} \theta \omega^b\right) = \sigma^a\,_b \omega^b + \frac{1}{2} \eta^{abcd} u_b \dot{u}_{c;d}, \tag2.5\\
& h^e\,_b \left(\omega^{bc} \, _{;c} - \sigma^{bc}\, _{;c} +  \frac{2}{3} \theta^{,b}\right) + (\omega^e\,_b + \sigma^e\,_b) \dot{u}^b = 0,\tag2.6\\
& \omega^a\,_{; a} = 2\omega^b \dot{u}_b,\tag2.7\\
& h_b\,\!^f \, h_b \,\!^g \dot{\sigma}_{fg}  - h_a\,\!^f \, h_b\,\! ^g \,\dot{u}_{(f;g)} - \dot{u}_a \dot{u}_b + \omega_a \omega_b + \sigma_{af} \sigma^f\,_b\\
& \qquad + \frac{2}{3} \theta \sigma_{ab} + h_{ab} \left(  - \frac{1}{3} \omega^2 - \frac{2}{3} \sigma^2 + \frac{1}{3} \dot{u}^c\,_{;c} \right) + E_{ab} =0,\tag2.8\\
& H_{ad} = 2 \dot{u}_{(a} \omega_ {d)}- h_a \,\!^ f \, h_d \,\! ^g (\omega_{(g} \, ^{b;c} + \sigma_{(g}\, ^{b;c}) \eta_{s)fbc} u^f,\tag2.9
\endalign$$
where $E_{ab}$ and $H_{ab}$ are, respectively, the electric and
magnetic parts of the Weyl tensor.$^\dag$

The remaining equations are the Bianchi identities 
(cf. Ellis, 1971, eqns. (4.21a -d)) and 
the field equations (cf. Ellis, 1971, eqns (4.23) - (4.26)).  In particular, in the case of zero vorticity ($\omega =0$) the Gauss-Codacci equations become
$$\align
^3\!R_{ab} & = (- \theta \sigma_{fg} - \dot{\sigma}_{fg} + \dot{u}_{(f;g)}) h_a \,\!^f \, h_b\,\!^g\\
& + \dot{u}_a \dot{u}_b + \frac{1}{3} h_{ab} \left(-\frac{2}{3} \theta^2 + 2\sigma^2 + 2 \mu - \dot{u}^c \, _{;c}\right),\tag 2.10
\endalign$$
and
$$^3\!R = -\frac{2}{3} \theta^2 + 2 \sigma^2 + 2 \mu, \tag 2.11$$
where $^3\!R_{ab}$ is the Ricci tensor of the $3$-spaces orthogonal to $u^a$ and $^3\!R$ is the corresponding Ricci scalar.

\subhead{B.  Integrability Conditions}\endsubhead

{}From equations (1.10) and (1.11), the existence of a kinematic 
self-similar vector implies that
$${\Cal L}_{\pmb{\xi}} g_{ab} = 2g_{ab} + 2(1 - \alpha) u_a u_b,  \tag 2.12$$
where $\alpha \neq 1$ is assumed hereafter.  From Yano (1955) we have that
$${\Cal L}_{\pmb{\xi}} R_{ab} =  ({\Cal L}_{\pmb{\xi}} \Gamma^c_{ab})_{;c} - ({\Cal L}_{\pmb{\xi}} \Gamma^c_{ac})_{;b}\tag2.13 $$  
where
$${\Cal L}_{\pmb{\xi}} \Gamma^a_{bc} \equiv \frac{1}{2} g^{at} [({\Cal L}_{\pmb{\xi}} g_{bt})_{;c} + ({\Cal L}_{\pmb{\xi}} g_{ct})_{;b} - ({\Cal L}_{\pmb{\xi}} g_{bc})_{;t}]. \tag2.14 $$
Hence we have that
$${\Cal L}_{\pmb{\xi}} R_{ab} = (1 - \alpha) g^{ct}[(u_bu_t)_{;ac} + (u_au_t)_{;bc}- (u_bu_a)_{;tc}].\tag2.15 $$
Therefore, using (2.1), after a long calculation we obtain
$$\split
\frac{1}{(1-\alpha)} {\Cal L}_{\pmb{\xi}} R_{ab} & =2 \dot{\sigma}_{ab} + 2 \theta \sigma_{ab} + 2 \sigma_{bc} \omega^c \,\! _a + 2 \sigma_{ac} \omega^c \,_b\\
& + 4 \omega_{ac} \omega^c\,\! _b + \frac{2}{3} \theta g_{ab} [\dot{\theta} + \theta] +
 u_a [2 \omega_{cb}\, ^{;c} - 2 \sigma_{bc} \dot{u}^c - 2 \omega_{bc} \dot{u}^c]\\
& + u_b [2 \omega_{ca} \, ^{;c} - 2 \sigma_{ac} \dot{u}^c - 2 \omega_{ac} \dot{u}^c] 
  + u_a u_b \left[ \frac{2}{3} \theta (\dot{\theta} + \theta) - 2 \dot{u}_c \, ^{;c} \right].
\endsplit \tag2.16 $$
Thus, decomposing equation (2.16) using $u^a$, we obtain
$$\align
& \left[\frac{1}{(1-\alpha)} {\Cal L}_{\pmb{\xi}} R_{ab}\right]u^a u^b = -8\omega^2 - 2 \dot{u}_c \,^{;c}, \tag2.17\\
& \left[\frac{1}{(1-\alpha)} {\Cal L}_{\pmb{\xi}} R_{ab}\right] h^{ab} = 2(\dot{\theta} + \theta^2 - 4 \omega^2), \tag2.18\\
& \left[ \frac{1}{(1 - \alpha)}{\Cal L}_{\pmb{\xi}} R_{ab} \right] u^a h^b_d   = 2 \omega_{da} \dot{u}^a + 2 \omega_{dc} \, ^{;c} - 4 \omega^2 u_d, \tag2.19\\
& \left[ \frac{1} {1-\alpha} {\Cal L}_{\pmb{\xi}} R_{ab} \right] \left[h^a_e h^b_f - \frac{1}{3} h_{ef} h^{ab}\right]  = 2 [\dot{\sigma}_{ef} - u_f \sigma_{eb} \dot{u}^b - u_e \sigma_{fb} \dot{u}^b\\
& \qquad \qquad \qquad + \theta \sigma_{ef} + \sigma_{fc} \omega^c\,\!_e + \sigma_{ec} \omega^c\,\!_f + 2 \omega_e \,^c \omega_{cf} + \frac{4}{3} h_{ef}\omega^2 ].
\tag2.20
\endalign $$

In the case of a perfect fluid, from the EFEs we have that
$$R_{ab} = \frac{1}{2} (\mu + 3p) u_a u_b + \frac{1}{2} (\mu -p) h_{ab}, \tag2.21 $$
whence, using equations (1.10) and (1.11), and (1.12), we obtain
$$\split
{\Cal L}_{\pmb{\xi}} R_{ab} & = \frac{1}{2}\{(a + 2 \alpha) \mu + 3(b + 2 \alpha) p\} u_au_b\\
& + \frac{1}{2} \{(a+2) \mu - (b+2) p\} h_{ab}. 
\endsplit \tag2.22 $$
Decomposing equation (2.22) yields
$$\align
\frac{1}{(1 - \alpha)} \left[{\Cal L}_{\pmb{\xi}} R_{ab}\right] u^a u^b  & = \left[ \frac{a + 2 \alpha}{2(1 - \alpha)}  \right] \mu + \frac{3[b+2 \alpha]}{2(1-\alpha)} p,\tag2.23\\
\frac{1}{(1-\alpha)} \left[ {\Cal L}_{\pmb{\xi}} R_{ab} \right]h^{ab} & = \left[ \frac{3(a+2)}{2(1-\alpha)} \right] \mu + \left[ \frac{-3(b+2)}{2(1-\alpha)} \right]p,\tag2.24\\
\left[ {\Cal L}_{\pmb{\xi}} R_{ab} \right] u^a h^b_d & = 0,  \tag2.25\\
\left[ {\Cal L}_{\pmb{\xi}} R_{ab} \right] \left( h^a_e h^b_f - \frac{1}{3} h_{ef} h^{ab} \right) & = 0. \tag2.26
\endalign 
$$
Hence, the integrability conditions for the existence of a proper kinematic self-similarity become
$$ \align
& -8\omega^2 - 2 \dot{u}_c \, ^{;c} = \left[ \frac{a + 2 \alpha}{2(1-\alpha)} \right] \mu + \left[ \frac{3[b + 2 \alpha]}{2(1 - \alpha)}  \right] p, \tag 2.27\\
& 2 (\dot{\theta} + \theta^2 - 4\omega^2) = \left[  \frac{3(a+2)}{2(1-\alpha)} \right] \mu + \left[\frac{-3 (b+2)}{2(1-\alpha)}  \right] p, \tag2.28\\
& 2 \omega_{da} \dot{u}^a + 2 \omega_{dc}\, ^{;c} - 4\omega^2 u_d =0,\tag2.29
\endalign $$
and 
$$\split
\dot{\sigma}_{ef} - u_f \sigma_{eb} \dot{u}^b -u_e \sigma_{fb} \dot{u}^b + \theta \sigma_{ef} + \sigma_{fc} \omega^c \, _e & + \\
   \sigma_{ec} \omega^c \,_f + 2 \omega_e\,\! ^c \, \omega_{cf} + \frac{4}{3} h_{ef} \omega^2 =& 0 . 
\endsplit \tag2.30$$ 
Equations (2.27)-(2.30) must be satisfied in addition to equations (2.2)-(2.9).

\subhead{C. Vorticity-free case}\endsubhead

In the Introduction we discussed the analogy with 
self-similarity in the Newtonian case where the irrotational case was singled out to be of special importance.  Setting
$\omega =0$ in the above equations we obtain:
$$\align
& \dot{\mu} + (\mu + p) \theta = 0, \tag2.31\\
& (\mu + p) \dot{u}_a = -p_{,b} h^b_a,\tag2.32\\
& \dot{\theta} + \frac{1}{3} \theta^2 - \dot{u}^a\,_{;a}  + 2 \sigma^2 + \frac{1}{2} (\mu + 3p) =0,\tag2.33
\endalign$$
and
$$\left(-\sigma^{bc} \,_{;c} + \frac{2}{3} \theta ^{,b}\right) h^e \,\!_b + \sigma^e \,\!_b \dot{u}^b =0.\tag2.34$$
The integrability conditions (2.27)-(2.30) become
$$\align
& - 2 \dot{u}_c \, ^{;c} = \left[\frac{a + 2 \alpha}{2(1 - \alpha)}\right] \mu + \left[\frac{3(b + 2 \alpha)}{2(1 - \alpha)}\right] p, \tag2.35\\
& 2 ( \dot{\theta} + \theta^2) = \left[ \frac{3(a+2)}{2(1 - \alpha)} \right] \mu + \left[ \frac{-3 (b +2)}{2(1 - \alpha)} \right]p, \tag2.36
\endalign $$
and
$$\dot{\sigma}_{ef} - u_f \sigma_{eb} \dot{u}^b - u_e \sigma_{fb}  \dot{u}^b + \theta \sigma_{ef} = 0.  \tag2.37 $$
Finally, equations (2.10) and (2.11) must be satisfied.  The remaining non-trivial equations are equations (2.8) and (2.9) 
which serve to define $E_{ab}$ and $H_{ab}$, respectively.
We note that from equations (2.8), (2.10), (2.11) and (2.37)  we have that
$$^3\!R_{ab} - \frac{1}{3} h_{ab}\, ^3\!R = E_{ab} + \sigma_{af} \sigma^f\,\!_b - \frac{1}{3} \theta \sigma_{ab} - \frac{2}{3} \theta^2 h_{ab}. \tag 2.38 $$

Now, contracting equation (2.37), we obtain 
$$\sigma[\dot{\sigma} + \theta \sigma] = 0. \tag2.39 $$
Using (2.35), equation (2.33) becomes
$$\dot{\theta} + \frac{1}{3} \theta^2 + 2 \sigma^2 + \frac{(a +2)}{4(1 - \alpha)} \mu + \frac{3(b +2)}{4(1 - \alpha)}p = 0. \tag2.40 $$
Differentiating this expression, and using equations (2.39) and (2.40), yields
$$\ddot{\theta} + \frac{8}{3} \theta \dot{\theta} + \frac{2}{3} \theta^3 + \frac{(a+2)}{4(1 - \alpha)} [\dot{\mu} + 2 \theta \mu] +  \frac{3(b+2)}{4(1 - \alpha)}[\dot{p} + 2 \theta p] = 0.\tag 2.41 $$
Finally, using equations (2.31) and (2.36), we obtain $(\theta \neq 0$)
$$\dot{\theta} + \theta^2 = \frac{3}{4} \frac{(a +2)}{(1 - \alpha)} (\mu -p). \tag2.42 $$
Hence from (2.36) we have that
$$(a-b) p =0. \tag2.43 $$
We shall study the special case of dust $(p =0)$ in the next 
subsection. If $p \neq 0$, then necessarily $a =b$.

\noindent
In either case, equations (2.40) and (2.42) yield
$$- \frac{2}{3} \theta^2 + 2 \sigma^2 + \frac{(a+2)}{(1 - \alpha)} \mu =0. \tag2.44 $$
{}From equation (2.11) we then have that
$$^3\!R = - \frac{(a + 2 \alpha)}{(1 - \alpha)} \mu. \tag2.45 $$

{}From the above equations we note a number of
interesting special cases arising.  When $p \neq 0$, we see that if
$p =\mu$ (stiff matter) or $a = -2$, equation (2.42) reduces to a simple first order DE for $\theta$.  In addition, if $a = -2$, then
from (2.44) we see that $\theta^2 = 3 \sigma^2$.  Finally, if $a = -2 \alpha$, then
from (2.45) we see that the Ricci scalar curvature of the $3$-spaces orthogonal to $u^a$ vanishes.

\heading{3. Special Cases}\endheading

\subhead {A. Geodesic case}\endsubhead

Let us first consider the case in which the
acceleration $\dot{u}_a$ is zero.  First, we note that when $u_a$ is 
irrotational and geodesic there exists preferred coordinates in which 
(Coley and McManus, 1994)
$$ds^2 = -dt^2 + H_{\alpha \beta} (t, x^\gamma) dx^\alpha dx^\beta, \tag3.1 $$
and 
$$u_a = -\delta^o_a \tag3.2 $$
(and in these coordinates we now have, for example, $\dot{\theta} = \theta_{, t}$).  We also note that if the shear is zero, $\sigma=0$, then the
spacetime is necessarily FRW (Ellis, 1971; Coley and McManus, 1994).  Henceforward we shall assume that $\sigma \neq 0$ (FRW spacetimes 
will be considered later).

Immediately, from (2.32) we have that
$$p_{,b} h^b\,_a = 0 \tag3.3 $$
(or $p_{, \alpha} =0$ in the preferred coordinates), and since
$$(a -b) p =0, \tag3.4 $$
equation (2.35) yields ($\mu + 3p \neq 0$)
$$a = -2 \alpha,  \tag3.5$$
whence (in both cases $a =b$ and $p =0$) it follows from equation (2.42) that 
$$\dot{\theta} + \theta^2 - \frac{3}{2} \mu + \frac{3}{2}p = 0, \tag3.6
$$
and consequently from (2.33) or (2.44) we obtain
$$-\frac{2}{3} \theta^2 + 2 \sigma^2 + 2 \mu =0. \tag 3.7 $$

Therefore, from (2.11) or (2.45), we find that
$$^3\!R = 0,  \tag3.8$$
and since, from equation (2.37), we have that
$$\dot{\sigma}_{ef} + \theta \sigma_{ef} = 0, \tag3.9 $$
and hence
$$\dot{\sigma} + \theta \sigma= 0, \tag 3.10 $$
from equation (2.10) we consequently have that
$$^3\!R_{ab} = 0; \tag 3.11 $$
i.e., the $3$-spaces orthogonal to $u^a$ are, in fact, Ricci flat.

Spacetimes in which the $3$-spaces orthogonal to $\bold{u}$ are Ricci
flat have been studied by a number of authors (e.g., 
Collins and Szafron, 1979; Stephani and Wolf, 1986; see also
Kramer et al., 1980).  In particular, examples in which
$\bold{u}$ is irrotational and geodesic include a subclass
of orthogonal spatially homogeneous models (i.e., the
Bianchi I spacetimes; see Kramer et al., 1980) and a 
subclass of the Szekeres cosmological models (Collins and Szafron, 1979;
Goode and Wainwright, 1982).  We shall return to this
and attempt to exploit previous work elsewhere.

The remaining non-trivial equations are [(2.31) and (2.34)]
$$\dot{\mu} + (\mu+p) \theta =0, \tag3.12 $$
and
$$(- \sigma^{bc}\,_{;c} + \frac{2}{3} \theta^{,b}) h^e\,\!_b = 0,   \tag3.13$$
which implies that
$$\sigma_e\,\!^c\,\!_{;c} = \frac{2}{3}\theta_{,e} + \left[2 \sigma^2 + \frac{2}{3} \dot{\theta}\right] u_e, \tag3.14 $$
and equations (2.8) and (2.9) yield simplified 
expressions for $E_{ab}$ and $H_{ab}$.
Finally, differentiating (3.6), and using equations (3.6) and (3.12), we obtain
$$\ddot{\theta} + 3 \theta \dot{\theta} + \theta^3 = -\frac{3}{2} (\dot{p} + 2p\theta). \tag3.15 $$

In the preferred coordinates (3.1), equations (1.10) and (3.2)  yield 
$$\xi^o\,\!_{,a} = \alpha \delta^o_a; \quad \xi^o = \alpha t + \bar{c}, \tag3.16 $$
and equations (1.12) and (3.3) yield
$$\frac{d}{dt} (p(t)) = \frac{bp(t)}{\alpha t + \overline{c}},  \tag3.17$$
the solution of which  can be written as $(\alpha \neq 0$\footnote"$^\dag$"{The special case $\alpha =0$ will be dealt with separately at the end of this subsection.})
$$p = p_o t^{-2}, \tag 3.18 $$
since from (3.5) either $b = -2 \alpha$ or $p =0$, and we have set $\bar{c} =0$ 
so that $p \rightarrow \infty$ as $t \rightarrow 0^+$.  In addition$^\dag$, equation (1.11) yields
$$\xi^\alpha\,\!_{,t} =0; \quad \xi^\alpha = \xi^\alpha(x^\gamma), \tag3.19 $$
and from equations (1.12) and (3.12) we obtain
$$\mu_{, \alpha} \xi^\alpha = -2 \alpha \mu + \alpha(p_o + \mu t^2) \theta t^{-1}. \tag3.20 $$

Next, the differential equation (3.15) becomes
$$\theta_{,tt} + 3 \theta \theta_{,t} + \theta^3 = 3p_o t^{-3} (1 - \theta t). \tag3.21 $$
Defining the new time, $\tau$, and the new variable, $\psi$, by 
$$\frac{d \tau}{dt} = \theta, \quad \psi = \theta^2, \tag3.22 $$
equation (3.21) becomes
$$\psi^{\prime \prime} + 3 \psi' + 2 \psi = 3p_ot^{-3}(\psi^{-1/2} - t), \tag3.23 $$
where $\psi' \equiv \frac{d \psi}{d \tau}$ and 
$$t = \int \psi^{-1/2} d \tau. \tag3.24$$

In principle we can solve equations (3.23) and (3.24) for 
$\psi = \psi (\tau, x^\gamma)$ and $\tau = \tau(t, x^\gamma$) in terms of arbitrary functions
of $x^\gamma$. We can
then integrate equation  (3.10) to obtain
$$\sigma^2 = \Sigma^2 (x^\gamma) e^{-2 \tau}. \tag3.25 $$
Equation (3.9) becomes
$$\sigma^\prime_{\alpha \beta}   - \frac{2}{\theta} \sigma_{\alpha \gamma} \sigma_\beta\,^\gamma + \frac{1}{3} \sigma_{\alpha \beta} = 0. \tag3.26 $$
In addition, equation (3.7) yields 
$$\mu = -\Sigma^2 e^{-2 \tau} + \frac{1}{3} \psi, \tag3.27 $$
where (eqn. (3.12))
$$\mu' + \mu = -p_o t^{-2}. \tag 3.28 $$

The remaining non-trivial equations serve to determine
the metric functions $H_{\alpha \beta}$ [e.g., eqns. (1.11) and (3.11)], or to 
constrain (through differential relationships) the various
arbitrary functions of $x^\gamma$ [e.g., eqns. (3.14), (3.20), (3.27) and (3.28)].  The special
case $p =0$ will be dealt with in the next subsection.  If $p \neq 0$ 
and there exists an equation of state of the form $p = p(\mu)$, 
then from equations (1.12) and (3.4) we obtain $p = (\gamma -1) \mu$ and 
hence
$$\mu = \mu_o t^{-2}. \tag 3.29 $$
{}From (3.12) we then obtain $(\mu_o + p_o \neq 0$) 
$$\theta = \frac{2 \mu_o}{\mu_o + p_o} t^{-1}, \tag3.30 $$
whence equation (3.6) yields
$$\mu_o = p_o \text{ or } p_o = \frac{2}{\sqrt{3}} \mu_o^{1/2} - \mu_o. \tag3.31 $$
In the latter case $\theta^2 = 3 \mu$ whence from (3.7) $\sigma =0$ and the 
spacetime is necessarily FRW, and in the former case we have
the special case of a stiff fluid with $\theta = t^{-1}$ and $\sigma^2= \frac{1}{3}t^{-2} (1 - 3 \mu_o)$ [and, of course,
$^3\!R_{ab} =0$ ; eqn. (3.20) is satisfied identically and $\sigma_{ab}$ 
satisfies eqns. (3.9) and (3.25), where $e^{- \tau} \equiv t^{-1}$, and $\sigma_\alpha\,\!^c \, \!_{;c} = 0$].
Finally, we note the {\it special non-trivial solution} of 
Benoit and Coley (1996) 
in the particular {\it case of spherical symmetry}.

\subhead{Special case $\alpha = 0$}\endsubhead 

In the special case of kinematic self-similarity of the 
zeroth kind (i.e., $\alpha =0$), the analysis up to equation (3.19) is similar resulting in
$$a = b =0, \tag3.32 $$
and 
$$p = p_o; \enskip p_o \text{ constant }, \tag3.33 $$
where 
$$\xi^a = (\bar{c}, \xi^\alpha(x^\gamma)). \tag3.34 $$
The differential equation (3.23), obtained from (3.15), becomes 
$$\psi^{\prime \prime} + 3\psi' + 2 \psi = -6p_o. \tag3.35 $$

This equation can be integrated to obtain
$$\theta^2 = \psi = c(x^\alpha) e^{-2\tau} + d(x^\alpha) e^{- \tau} - 3p_o, \tag3.36 $$
and the analysis then essentially follows the same steps as in the zero-pressure case (see the details in the next subsection).
Indeed, integrating equations (3.12) and (3.10) to obtain
$$\mu = M(x^\alpha) e^{-\tau} - p_o, \tag3.37 $$
and
$$\sigma^2 = \Sigma^2 (x^\alpha) e^{-2\tau} \tag3.38  $$
we can see that equations (3.6), (3.7) [(3.9) and (3.10)], and
(1.12) [the analogue of (3.20)] are all invariant under $\theta \rightarrow \bar{\theta}$,
$\mu \rightarrow \bar{\mu}, \sigma \rightarrow \bar{\sigma}$ where $\bar{\theta}^2 = \theta^2 + 3p_o$, $\bar{\mu} = \mu + p_o, \bar{\sigma}^2 = \sigma^2$.
However, assuming that $p_o >0$ (i.e., positive pressure), we see from
(3.36) that the model is only valid for $\tau \leq \tau_c$, for some
critical value $\tau_c$ (so that $\theta^2 \geq 0$), and for early times
($\tau \rightarrow - \infty$ or $t \rightarrow 0^+$) the model is indistinguishable from the 
zero-pressure model.

\subhead{B. Pressure-free case}\endsubhead

The special case of dust is of particular importance
since in this case the constraints from the imposition of 
self-similarity are the least restrictive and also because there are
examples of dust models admitting a kinematic
self-similarity (Carter and Henriksen, 1989).  We also note
that in the case of dust $(p =0)$, it follows that 
${\Cal L}_\xi T_{ab} = (a + 2\alpha) T_{ab}$; that 
is the \underbar{total} energy-momentum tensor satisfies an equation of the form (1.12).  

When $p=0$, the differential equation (3.23) reduces
to
$$\psi^{\prime \prime} + 3 \psi' + 2 \psi =0, \tag3.39 $$
with the solution
$$\psi = \theta^2 = c(x^\alpha) e^{-2 \tau} + d(x^\alpha) e^{-\tau}, \tag3.40 $$
where $(cd \neq 0)$
$$t = \frac{2 \sqrt{c}}{d} \left(1 + \frac{d}{c}e^\tau\right)^{1/2} - \frac{2 \sqrt{c}}{d} \tag3.41 $$
(where the function of integration is chosen to ensure that
$t \rightarrow 0^+$ as $\tau \rightarrow - \infty$).  Alternatively, we can write 
$$\theta^2 = 4 \left(t+\frac{2 \sqrt{c}}{d}\right)^2 \left[ \frac{4c}{d^2} - \left( t + \frac{2 \sqrt{c}}{d} \right)^2  \right]^{-2}, \tag3.42 $$
where 
$$e^\tau = \frac{d}{4} \left(t + \frac{2 \sqrt{c}}{d}  \right)^2 - \frac{c}{d}. \tag 3.43 $$

Integrating equation (3.28) yields
$$\mu  = M(x^\alpha) e^{-\tau},\tag3.44 $$
and equation  (3.25) gives
$$ \sigma^2  = \Sigma^2 (x^\alpha) e^{-2 \tau},\tag3.45$$
whence equations (3.6) and (3.7) (or (3.27)) are satisfied when
$$\align
M & = \frac{1}{3}d, \tag 3.46\\
\Sigma^2 & = \frac{1}{3}c.  \tag 3.47
\endalign $$
In addition, since $a = -2 \alpha \, (\mu \neq 0)$, equation (1.12) yields
$$2 \alpha + \frac{M_{, \alpha} \xi^\alpha}{M} - \tau_{, \alpha} \xi^\alpha - \theta \xi^o = 0, \tag3.48 $$
where 
$$\xi^a = (\alpha t + \bar{c}, \xi^\alpha(x^\gamma)). \tag3.49 $$
Using equations (3.42), (3.43) and (3.46), equation (3.48) reduces to 
$$\left\{\left(\frac{\sqrt{c}}{d}\right)_{,\alpha} \xi^\alpha- \frac{\alpha \sqrt{c}}{d}\right\} t + \left[t + \frac{2 \sqrt{c}}{d}\right] \frac{\bar{c}}{2} = 0, \tag3.50$$
and hence $\bar{c} =0$ (and $\alpha$ is necessarily non-zero) and
$$\left(\frac{\sqrt{c}}{d}\right)_{, \alpha} \xi^\alpha = \alpha \frac{\sqrt{c}}{d}. \tag3.51$$
Equations (3.14), (1.11), (3.9) and (3.11), remain to be satisfied; these
equations (through equation (3.1) and due to the
definitions of the shear and expansion in terms of 
the metric functions, their time derivatives, and the inverse 
metric functions), constrain the metric functions $H_{\alpha \beta}$.  We recall that
the resulting spacetimes are $3$-Ricci flat (i.e., $^3\!R_{ab} =0$).

Let us investigate the asymptotic 
behaviour of the solutions.  As $t \rightarrow \infty$ ($\tau \rightarrow \infty$), we find that
$$  \theta = \sqrt{d} e^{-\tau/2} = 2t^{-1},   \tag 3.52 $$
and from equations (3.46) and (3.47) we have that
$$ \frac{\mu}{\theta^2} \rightarrow \frac{1}{3} \text{ and } \frac{\sigma}{\theta} \rightarrow 0.  \tag 3.53$$  
Therefore, the models are asymptotic (at late times) to an exact
zero-pressure and flat FRW model with a power-law scale 
function; we note that this  Einstein de Sitter model admits a homothetic vector.

On the other hand, as $t \rightarrow 0^+$ ($\tau \rightarrow - \infty$), we find that
$$\theta = \sqrt{c} e^{-\tau} = \frac{1}{t}, \tag3.54 $$
and from equations (3.46) and (3.47) we have that
$$\frac{\mu}{\theta^2}
\rightarrow 0 \text{ and } \frac{\sigma^2}{\theta^2} \rightarrow \frac{1}{3}. \tag3.55 $$
Hence the models are asymptotic (at early times) to an exact
vacuum, $3$-Ricci flat solution; in particular, this 
exact solution has
$$\leqalignno{ \mu &= 0,\cr
\theta = \frac{1}{t}; \quad & \sigma^2 = \frac{1}{3} \theta^2 = \frac{1}{3} t^{-2},\cr
^3\!R_{ab} & = 0.\cr}$$
{}From equation (3.1) and from the definitions of the shear
and expansion, these equations yield 
the following equations for the metric functions $H_{\alpha \beta} (t_, x^\gamma)$:
$$\align
H^{\alpha \beta} H_{\alpha \beta , t} & = \frac{2}{t}, \tag3.56\\
H^{\alpha \gamma} H^{\beta \delta} H_{\alpha \beta , t} H_{\gamma \delta, t} & = \frac{4}{t^2}, \tag3.57
\endalign $$
and (from equation (3.9))
$$H_{\alpha \beta, tt} - H^{\gamma \delta} H_{\alpha \gamma, t} H_{\beta \delta, t} + \frac{1}{t} H_{\alpha \beta, t} = 0. \tag3.58 $$
In the special case of spatial homogeneity we 
necessarily obtain the Kasner model.  In particular,
in general in the case that $H_{\alpha \beta} (t_, x^\gamma)$ is diagonal, i.e., $H_{\alpha \beta} \equiv \text{ diag } \{h_1 (t_, x^\gamma)$, $h_2(t_, x^\gamma)$, $h_3(t_, x^\gamma) \}$, these equations can be
integrated to yield
$$h_1  = A_1(x^\gamma) t^{2a_1}, h_2 = A_2(x^\gamma) t^{2a_2}, h_3 = A_3(x^\gamma) t^{2a_3},  \tag3.59 $$
where the constants in (3.59) obey
$$a_1 + a_2 + a_3 = a_1\,^2 + a_2\,^2 + a_3\,^2 = 1, \tag3.60$$
and hence we obtain the Kasner model.  We note that
this vacuum Bianchi I exact solution also admits a homothetic vector.

Finally, we note that the spherically symmetric dust solutions 
of Lynden-Bell and Lemos (1988) and Carter and 
Henriksen (1989), that are a particular case of the perfect
fluid solutions of Benoit and Coley (1996), represent a special
non-trivial solution of the above equations.

\subhead{C. Case of Zero Expansion}\endsubhead

We shall also consider the special case in which $\theta =0$, although
this case is not of interest from a cosmological point of view.  Let us
choose comoving coordinates so that (with $\omega =0$)
$$ds^2 = -(U^2) dt^2 + H_{\alpha \beta} (t_, x^\gamma) dx^\alpha dx^\beta, \tag3.61 $$
where 
$$u^a = \frac{1}{U} \delta^a_0. \tag3.62 $$
Using these coordinates we have that
$$\dot{u}_a = (0, [ln U]_{, \alpha}), \tag3.63$$
$\theta =0$ implies that 
$$H^{\alpha \beta} H_{\alpha \beta, t} =0, \tag3.64 $$
and $\dot{\Phi} = 0$ implies that $\Phi_{, t}=0$.  In addition, from equations
(1.10) and (1.11) we obtain
$$\xi^a = (\xi^0(t), \xi^\alpha(x^\gamma)).\tag3.65$$
{}From equations (2.31) and (2.39) we have that
$$\mu_{, t} = 0 = \sigma_{, t}, \tag3.66 $$
and from equation (3.41) we find that
$$(\mu + p) [ln U]_{, \alpha} = - p_{, \alpha}.  \tag3.67 $$
The class of \underbar{static solutions} is contained
within this special case presently under consideration.

Immediately, equation (2.36) yields
$$(a+2) \mu - (b+2) p =0, \tag3.68 $$
whence equations (2.33) and (2.35) yield
$$8(1- \alpha) \sigma^2 + (a+2) \mu + 3(b+2) p=0. \tag3.69 $$
These two equations are best dealt with separately in two 
different subcases.

\noindent
{\bf Subcase (i)}:  Either $b = -2$ or $p=0$.  From equation (3.68) $a = -2$,
and hence from  equation (3.69), the shear is zero, $\sigma^2 =0$.
Neglecting the case $p =0$ (since in this case we obtain
$\dot{u}_a =0$ from eqn. (2.32) and hence the resulting spacetime is a special
static FRW spacetime), we consequently obtain
$$a = b = -2, \tag3.70$$
and hence
$$\sigma_{ab} =0; \tag3.71 $$
i.e.,
$$H_{\alpha \beta, t} = 0  \tag3.72$$
[and consequently $H_{\alpha \beta}$ can be diagonalized$-H_{\alpha \beta} = \text{ diag } \{h_1(x^\gamma), h_2(x^\gamma), h_3(x^\gamma)\}]$.   
{}From equations (2.11) and (2.33) we then obtain
$$^3\!R = 2 \mu(x^\gamma), \tag3.73 $$
and 
$$\dot{u}^a\,_{;a} = \frac{1}{2} (\mu + 3p).  \tag3.74 $$
In addition, from equation (2.9), we have that
$$H_{ab} = 0, \tag3.75 $$
i.e., the magnetic part of the Weyl tensor vanishes (note clumsy notation here), 
and equation (2.38) yields
$$E_{ab} = ^3\!\!R_{ab} - \frac{1}{3} \,^3\!Rh_{ab}. \tag3.76$$
{}From equation (1.10) we have that
$$(ln U)_{,t} \xi^0 + (ln U)_{, \alpha} \xi^\alpha = \alpha - \xi^0_{,t}, \tag3.77 $$
and equations (1.12) yield
$$\align
\mu_{, \alpha} \xi^\alpha & = -2 \mu,\tag3.78 \\
p_{, t} \xi^0 + p_{, \alpha} \xi^\alpha & = -2p. \tag3.79
\endalign $$   
Finally, equation (3.67) gives
$$(\mu + p) (ln U)_{, \alpha} \xi^\alpha = -p_{, \alpha} \xi^\alpha, \tag3.80$$
and (the integrability conditions of (3.67) give)
$$p_{, \alpha} \mu_{, \beta} = \mu_{, \alpha} p_{, \beta}.  \tag3.81 $$

{}From this equation we obtain the solution
$$ln p = F(t, ln \mu), \tag3.82$$
whence on defining $T$ by
$$\frac{dT}{d t} = - \frac{2}{\xi^0}, \tag3.83$$
equation (3.79) becomes
$$\frac{\partial F}{\partial (ln \mu)} + \frac{\partial F}{\partial T} = 1, \tag3.84$$
with the solution
$$F = \frac{1}{2} (ln \mu + T) + \bar{f}(ln \mu - T), $$
which can be written as
$$p = \mu f(ln \mu - T),  \tag3.85 $$
where $f$ is an arbitrary function of a single variable.  Equations
(3.67), (3.77) and (3.80) then yield
$$(ln U)_{, T} = \frac{1}{2} (\xi^0_{, t} - \alpha) + \left[\frac{f+f'}{1+f}  \right] $$
and
$$(ln U)_{, \alpha} = - \left[ \frac{f+f'}{1+f} \right] (ln \mu)_{, \alpha} $$
which has the solution
$$ln U = \int \left[ \frac{\alpha - \xi^0_{,t}}{\xi^0}\right]d t - \int \frac{f(m)+f'(m)}{1 +f(m)}dm ,  \tag3.86 $$
where $m \equiv (ln \mu -T)$.

Now,
$$\dot{u}^a \, _{;a} = H^{\alpha \beta} (x^\delta) [(ln U)_{, \alpha \beta} + (ln U)_{, \alpha} (ln U)_{, \beta} - \Gamma^\delta_{\alpha \beta} (x^\delta)(ln U)_{, \gamma}], \tag3.87 $$
whence, on defining
$${\Cal F}(m) = \int \frac{f(m) +f'(m)}{1+f(m)}dm, \tag3.88 $$
equation (3.74) becomes
$$\split
& [{\Cal F}']\{H^{\alpha \beta}(ln \mu)_{, \alpha \beta} - H^{\alpha \beta} \Gamma^\gamma  _{\alpha \beta} (ln \mu)_{, \gamma}\} \\
& \qquad + [{\Cal F}^{\prime \prime} - ({\Cal F}')^2] \{H^{\alpha \beta}(ln \mu)_{, \alpha}(ln \mu)_{, \beta}\}\\
&\qquad \quad + \left[  \frac{1}{2}(1+3f)\right] \{\mu\}=0. 
\endsplit \tag3.89 $$
This equation contrains the metric functions $H_{\alpha \beta} \,  (x^\gamma)$ and
$\mu(x^\gamma)$
[and $f(ln \mu - T(t))\, ({\Cal F})$ and $\xi^0(t)$].  The $H_{\alpha \beta}$ 
[and $\xi^\alpha(x^\gamma)$] are further
constrained by equations
(3.73), (3.75) and (3.76) [and eqns. (3.78) and (1.11)], where $\mu$ is given in terms of the metric functions through the EFEs.

We notice that equation (3.89) is of the form of the sum 
of (three) terms in which each term in square brackets depends on $t$ [through eqns. (3.65), (3.83), (3.85) and (3.88)] and 
each term in curly brackets does not.  A particular solution of
equation (3.89) has \underbar{$f=$ constant}.   Indeed, if $f$ is not constant then 
 {\it in general} (for $\mu \neq 0$) we have that
$${\Cal F}^{\prime \prime} - ({\Cal F}')^2 = c_1 (1 + 3f); \, {\Cal F}' = c_2 (1 + 3f)\, [c_2 \neq 0],  \tag3.90$$
which on integration yields $e^{-{\Cal F}} = d_2 - \frac{c_2}{c_1}d_1 e^{c_1m/c_2} (c_1 \neq 0)$ or
$e^{-{\Cal F}} = d_2 - d_1m \,(c_1 =0)$ and hence a particular functional form for $f(m)$, whence for consistency
equation (3.88)
implies that $d_1 =0$ (in either case), which  {\it contradicts} the  {\it assertion} that $f$
is not constant.

Therefore, in general $f$ is constant and hence
from (3.85) we have that
$$p = (\gamma -1) \mu; \, f \equiv \gamma-1\, (\text{const}.). \tag3.91$$
In particular, we note that $p = p(x^\gamma)$.  Finally, equation (3.67) gives
$$[ln U]_{, \alpha} = \left( \frac{1 - \gamma}{\gamma} \right) (ln \mu)_{, \alpha}, 
\tag3.92$$
which integrates to yield
$$U = u(t) \mu^{-1+1/\gamma}. \tag3.93$$
However, since $U$ is separable, a redefinition of the time
coordinate in (3.61) can be employed to set $u(t) =1$.  
Hence
$$U = U(x^\gamma) = \mu^{-1+1/\gamma},  \tag3.94 $$
and the metric is independent of time and hence the
model is {\it completely static}.

Equation (3.77) then yields
$$\xi^0 (t) = \left(\alpha + \frac{2(1 - \gamma)}{\gamma}  \right) t + \overline{c}, \tag 3.95$$
whence the $\xi^\alpha(x^\gamma)$ are constrained by equations (3.78) and
(1.11) while the $H_{\alpha \beta}(x^\gamma)$ are themselves subject to
equations (3.73), (3.75) and (3.76).

Finally, from (3.91) we have that
$${\Cal F}(m) = \int \frac{\gamma -1}{\gamma} dm = \frac{\gamma -1}{\gamma} (ln \mu - T),\tag3.96$$
and equation (3.86) yields
$$\split
ln U & = - \frac{1}{2}  \int (\alpha - \xi^0_{,t}) dT - {\Cal F}(m)\\
& = \int \frac{(1 - \gamma)}{\gamma} dT - \frac{\gamma -1}{\gamma} (ln \mu - T) = \frac{1 - \gamma}{\gamma} ln \mu + d, 
\endsplit  \tag3.97 $$
which is consistent with equation (3.93).  In addition, in this case
equation (3.89) reduces to
$$\split
& \frac{(\gamma -1)}{\gamma} H^{\alpha \beta} (ln \mu)_{, \alpha}(ln \mu)_{, \beta} - H^{\alpha \beta} (ln \mu)_{, \alpha \beta}\\   
&  \qquad + H^{\alpha \beta} \Gamma^\gamma_{\alpha \beta} (ln \mu)_{, \gamma} = \frac{(3 \gamma -2) \gamma}{2(\gamma -1)} \mu. 
\endsplit \tag3.98 $$
This equation must be satisfied in addition to equations (3.73), (3.75) and (3.76).

\noindent
{\bf Subcase (ii)}:  In this subcase we have that
$b \neq -2$ and $p \neq 0$, so that equation (3.68) yields
$p = \frac{(a+2)}{(b+2)} \mu$, whence equations (1.12) 
imply that $a= b \,(\neq -2)$ and 
hence $p = \mu$ (note that $p_{,t} =0$), and finally equation (3.69)
yields $2(1 - \alpha)\sigma^2 + (a+2) \mu =0$; that is, we have that
$$(b+2) p \neq 0, a =b, p = \mu, \sigma^2 = \frac{-(a+2)}{2(1 -\alpha)} \mu. \tag3.99 $$
{}From equations (2.11) and (2.33) we also obtain
$$^3\!R = \frac{-(a+2 \alpha)}{(1 - \alpha)} \mu = \dot{u}^a \, _{;a}. \tag3.100$$

{}From equation (3.67) we obtain
$$U = u(t) \mu^{-\frac{1}{2}} \tag3.101$$
(which we note is the special case of (3.93) with $\gamma =2 \, -$ corresponding 
to $p = \mu$), whence by redefining the $t$-coordinates to set $u(t) =1$
we have that
$$U = \mu^{-\frac{1}{2}}. \tag3.102$$
Therefore, although $\mu, p$ and $U$ are independent of $t$,
the metric functions $H_{\alpha \beta} = H_{\alpha \beta} (t_, x^\gamma)$ depend on $t$ and 
the models are not static (since $\sigma^2 \neq 0$ for 
$\mu \neq 0$, $\sigma_{ab} \neq 0$ and hence $H_{\alpha \beta, t}$ cannot vanish for all $\alpha_, \beta$).
However, the $H_{\alpha \beta}$ are constrained by equations (3.64), 
(3.99) and (3.100), in addition to equations
(2.34) and (2.37), etc.  For example, using (3.63) and (3.64), equation (3.100) yields
$$U[U_{, \alpha \beta} - \Gamma^\delta _{\alpha \beta} U_{, \delta}] H^{\alpha \beta} = \frac{-(a+2\alpha)}{(1 - \alpha)} \enskip (\text{const.}). \tag3.103$$                                                                                

Using equations (3.99) and (3.102), equations (1.12) yield
$$(ln U)_{, \alpha} \xi^\alpha = -\frac{a}{2} = -\frac{1}{2} (ln \mu)_{, \alpha} \xi^\alpha, \tag3.104 $$
whence equation (1.10) yields
$$\xi^0_{, t} = \alpha + \frac{a}{2}, \tag3.105 $$
which integrates to give 
$$\xi^0 (t) = \frac{1}{2} (a+2 \alpha) t + \overline{c}. \tag3.106 $$
 Equation (1.11) remains to be satisfied.

\heading{4. More Special Cases}\endheading

There are two special cases of particular interest in which 
the kinematic self-similarity is either parallel or orthogonal to the velocity vector. Let us next consider these two cases separately.

\subhead {A.  $\pmb{\xi}$ parallel to $\bold u$}\endsubhead

In Coley (1991) it was shown that if a perfect fluid spacetime 
admits a proper HV parallel to the velocity vector then that spacetime 
is necessarily an FRW spacetimes satisfying $\dot{\theta} = 
-\frac{1}{3} \theta^2, \theta \neq 0$  
[with the very special and physically unreasonable 
equation of state $\mu + 3p =0$;  strictly 
speaking this result was only proven in the case
$p = p(\mu)$ and $\mu + p \neq 0$].

Suppose that a spacetime admits a kinematic self-similar vector 
$\pmb{\xi}$ parallel to ${\bold u}$, i.e.,
$$\xi^a = Au^a, \tag4.1 $$
then, from equation (2.12),
$$A(u_{a;b} + u_{b;a}) + A_{, b} u_a + A_{,a} u_b = 2g_{ab} + 2(1 - \alpha) u_a u_b. \tag4.2$$
Contracting equation (4.2) with $g^{ab}$ and $u^a u^b$ in turn then yields
$$\dot{A} = \alpha, \, A \theta=3, \tag4.3 $$
and hence
$$\dot{\theta} + \frac{\alpha}{3} \theta^2 = 0. \tag4.4 $$
Contracting equation (4.2) with $h_c \,^a h_b\,\!^d$ then yields
$$\sigma_{ab} = 0, \tag4.5 $$
i.e., the spacetime is shear-free, and finally contracting equation (4.2) with $u^a$ yields
$$\dot{u}_a = (ln A^{-1})_{,b} h^b_a. \tag4.6 $$
Assuming that the vorticity is zero, equation (2.6) then yields
$$\theta_{,b}h^b_c =0. \tag4.7 $$
Therefore, from equations (4.3) $[A = 3/\theta]$ and (4.7) we see that
$$\dot{u}_a =0, \tag4.8 $$
i.e., the acceleration is zero.

Hence, since the shear, vorticity and acceleration are all 
zero, the spacetime is necessarily FRW (Ellis, 1971; Coley and 
McManus, 1994).  Therefore, there exists coordinates in which the metric is of the form (4.48) and equation (4.4) yields
$$\theta_t + \frac{\alpha}{3} \theta^2 = 0; \enskip \theta = \frac{3}{\alpha} t^{-1} \tag4.9 $$
(defining the constant of integration so that $\theta \rightarrow \infty$ as $t \rightarrow 0^+$).  The remaining equations to be solved reduce to
$$\align
\frac{a + 2 \alpha}{2(1- \alpha)} \mu & + \frac{3(b + 2 \alpha)}{2(1 - \alpha)}p = 0,\tag4.10\\
2 \theta^2 \left(1 - \frac{\alpha}{3} \right) & = \frac{3(a+2)}{2(1 - \alpha)} \mu - \frac{3(b+2)}{2(1 - \alpha)} p,\tag4.11\\
2 \theta^2(1 - \alpha) & = -3 \mu - 9p, \tag4.12
\endalign
$$
and 
$$\dot{\mu} + (\mu + p) \theta = 0. \tag4.13 $$
It can be easily shown that $a \neq  -2 \alpha$ and $p =0$ lead to contradictions, so that (4.10) yields
$$a = b = -2 \alpha. \tag4.14$$
Equations (4.11)-(4.13) then yield the consistent solution
$$p = \left[\frac{2 \alpha}{3} -1 \right] \mu\tag4.15 $$
and 
$$\mu = \frac{1}{3} \theta^2 = \frac{3}{\alpha^2} t^{-2}. \tag4.16 $$

We note that the above solution reduces to the homothetic case 
when $\alpha = 1$ (see, e.g., eqns. (4.4) and (4.15) and (4.16)).  However, unlike the 
homothetic case there exist models with realistic equations of state; 
for example, $0 < p < \mu$ for $\frac{3}{2} < \alpha < 3$.

\subhead{B. $\pmb{\xi}$ orthogonal to $\bold u$}\endsubhead

McIntosh (1975) showed that a perfect fluid spacetime
cannot admit a non-trivial homothetic vector which is orthogonal to the 
fluid $4$-velocity unless $p = \mu$.

We shall work in comoving coordinates in which the
metric is given by (3.61) and the velocity vector by (3.62). In these
coordinates the acceleration is then given by
$$\dot{u}_0 =0; \dot{u}_\alpha = (ln U)_{, \alpha}. \tag4.17$$
Since $\pmb{\xi}$ is orthogonal to $\bold u$, $\xi^0=0$, and equation (1.11) implies that $\xi^\alpha = \xi^\alpha(x^\gamma)$, and consequently equation (1.10) reduces to
$$U_{, \alpha} \xi^\alpha = \alpha U.  \tag4.18 $$
Equations (1.12) yield
$$\mu_{, \alpha} \xi^\alpha = a \mu \tag4.19$$
and
$$p_{, \alpha} \xi^\alpha = b p. \tag4.20$$
Also, we can further specify the coordinates so that
$$\xi^\alpha = \xi(x^\gamma) \delta^\alpha_x, \tag4.21$$
and in these coordinates (if $\alpha ab \neq 0$) equations (4.18) - (4.20) yield
$$(ln [U^{1/\alpha}])_{, x} = (ln [\mu^{1/a}])_{,x} = (ln[p^{1/b}])_{,x} = \frac{1}{\xi}, \tag4.22 $$
which can be partially integrated to yield
$$\mu = \overline{f}(t,y,z) U^{a/\alpha}; p = \overline{g}(t,y,z) U^{b/\alpha}.  \tag4.23 $$

Finally, the conservation equation (2.32) reduces to
$$(\mu +p) (ln U)_{, \alpha} = -p_{, \alpha}, \tag4.24$$
whence, on contraction with $\xi^\alpha$ and using equations (4.18) and (4.20), we obtain
$$\alpha [ \mu + p] = -bp. \tag4.25 $$
Apart from the special subcase $\alpha =b =0$, which implies that
$$U_{, x} =0 = p_{, x}, \tag4.26$$
equation (4.25) implies that (i) if $b =0$, then either $\alpha =0$ 
(the special subcase above) or $\mu +p =0$ (a case which we shall not
consider here), (ii) if $\alpha =0$, then either $b =0$ (special subcase)
or $p =0$ (a case considered earlier), and (iii) if $b = - \alpha$, then
either $\alpha =0$ (and $b =0$) or $\mu =0$ (a case of no interest here; however, see the appendix), otherwise we have that in the 
\underbar{general} case
$$p = \frac{- \alpha}{(\alpha +b)} \mu; \enskip \alpha b(\alpha +b) \neq 0. \tag4.27 $$
Therefore, equations (4.27) and (1.12) immediately imply that
$$a =b. \tag4.28$$
The conservation equation (2.31) then yields
$$\mu_{, t} = -a(a + \alpha)^{-1} \mu U\theta, \tag4.29$$
which can be regarded as an equation for $\theta$ in terms of 
$\mu$ and $U$.

Using equations (4.27) and (4.28), equations (2.33) and 
(2.35) then yield
$$\dot{\theta} + \frac{1}{3} \theta^2 = -2\sigma^2 - \frac{(a+2)(a-2\alpha)}{4(a+ \alpha)(1 - \alpha)} \mu,  \tag4.30 $$
equation (2.36) yields
$$\dot{\theta} + \theta^2 = \frac{3(a+2)(a+2 \alpha)}{4(1-\alpha)(a+\alpha)} \mu, \tag 4.31$$
from which we deduce that
$$\frac{2}{3} \theta^2 = 2 \sigma^2 + \frac{(a+2)}{(1-\alpha)} \mu. \tag 4.32$$
Differentiating this equation with respect to $t$, and using
equations (4.29) and (4.31), we obtain $(\sigma \neq 0)$
$$\dot{\sigma} = -\sigma \theta, \tag4.33$$
which also follows from equation (2.37).  Finally, from equations
(4.32) and (2.45) we obtain
$$^3\!R = - \frac{(a+2\alpha)}{(1 - \alpha)} \mu. \tag4.34$$

Now, defining $F(x,y,z)$ by $(ln F)_{, x} = 1/\xi$, 
equations (4.22) then yield
$$\align
\mu & = f(t,y,z) F^a, \tag4.35\\
U & = g(t,y,z) F^\alpha . \tag4.36
\endalign $$
Since $F_{, t} = 0$, equation (4.29) yields 
$$\theta = -\frac{(a+\alpha)}{a} \frac{f_{, t}}{fg} F^{- \alpha}, \tag4.37 $$
whence equation (4.31) then yields
$$- \left[\frac{f_{,t}}{fg}\right]_{,t} + \frac{(a+\alpha)}{a}\, \frac{(f_{, t})^2}{f^2 g} = \frac{3a(a+2)(a+2 \alpha)}{4(1 - \alpha)(a+\alpha)^2} fg F^{a + 2 \alpha}.  \tag4.38$$
Because the only term in (4.38) that depends on $x$ is the term
$F^{a + 2\alpha}$ (and since $F_{, x} \neq 0$ from $(ln F)_{,x} = 1/\xi \neq 0)$,
and assuming that $a \neq 0$ (since $a =0$ implies $b =0$), the
only way that this equation can then be satisfied is for either 
$$ (i) \enskip   a = -2, \text{ or (ii) } a = -2 \alpha.  \tag4.39$$
Hence, from equation (4.31) we have that
$$\dot{\theta} + \theta^2 =0. \tag4.40 $$
{}From equations (4.38) and (4.40) we can then integrate to obtain
$$g = J(y,z) f^{-(a+2 \alpha)/\alpha} f_{, t}. \tag4.41$$

Finally, let us consider the two cases in (4.39) separately.

\noindent
{\bf (i)  $a = -2$}.  In this case (4.25) implies that
$$p = \frac{\alpha}{2 - \alpha} \mu, \tag4.42$$
and equations (4.32) and (4.34) imply that
$$\sigma^2 = \frac{1}{3} \theta^2, \tag4.43 $$
and 
$$^3 \!R = 2 \mu. \tag4.44  $$

\noindent
{\bf (ii) $a = - 2\alpha$}.  In this case (4.25) implies that
$$p = \mu  \tag4.45 $$
(i.e., the fluid is stiff), and from equations (4.32) and (4.34) we have that
$$\mu + \sigma^2 = \frac{1}{3} \theta^2, \tag4.46$$
and hence 
$$^3\!R = 0. \tag4.47$$

Further progress (e.g., further constraining the form of
the functions $F(x,y,z), f(t,y,z)$ and $J(y,z)$ or determining the
form of the metric functions $H_{\alpha \beta} (t,x,y,z)$) can be made in 
specific spacetimes.  For example, spherically symmetric spacetimes have been
studied [and it has been claimed that in such spacetimes if there exists a vector field $\pmb{\xi}$ 
satisfying (1.11) which is orthogonal to $\bold{u}$, then the resulting spacetime metric is singular (Ponce de Leon, 1993)].  We shall investigate the existence of kinematic self-similar vectors in
the special case of FRW spacetimes in the next subsection.

\subhead{C. FRW models}\endsubhead

We recall from Maartens and Maharaj (1986) that (perfect fluid) FRW models, when written in the comoving form
$$ds^2 = -dt^2 + R^2(t) \left\{ \frac{dr^2}{1 -kr^2} + r^2 d \Omega^2 \right\}, \tag4.48 $$
admit a homothetic vector $\bold{P}_o ( = \pmb{\partial}_t)$ parallel to $\bold{u}$ for all $k$ when 
$R(t) =dt$, $d$ constant (and hence $\mu = -3p =3(1 + kd^{-2})t^{-2}$; $\mu + 3p = 0$), and 
admit a homothetic vector ${\bold H} \, ( = t \pmb{\partial}_t + r \pmb{\partial}_r$) when $R = dt^c$ but only in 
the case $k =0$ (whence $\mu = 3c^2 t^{-2}$, $p = (\gamma -1) \mu$ with $\gamma = 2/3 c$).

We cannot simply apply the results from section 3 here since it was 
explicitly assumed there that 
$\sigma \neq 0$.  However, when $\sigma = \omega = \dot{u}_a =0$, we immediately obtain from equation (2.35) that
$$(a + 2 \alpha) \mu + 3 (b + 2 \alpha) p = 0.  \tag4.49$$
This again suggests two subcases.  First, when $a = -2\alpha$, we have
that either $b = -2 \alpha$ or $p =0$, whence from equations (2.33) and 
(2.36) we obtain (in either case) $\mu = \frac{1}{3} \theta^2$ and hence from
equation (2.11) we obtain $^3\!R =0$.  Second, if $a \neq -2 \alpha$, then if
$b = a$, from (4.49) we obtain $\mu +3p =0$ [on the other hand, if we assume $a \neq b \neq -2 \alpha \, (p \neq 0)$, then from (4.49) we have that $p = [-(a+ 2 \alpha)/3(b + 2 \alpha)] \mu$, whence
from equations (1.12) we deduce that $a =b$, resulting in a 
contradiction].  Therefore, in order for an FRW spacetime to admit
a kinematic self-similarity then necessarily either $k =0$ (zero-curvature) or 
$\mu + 3p =0$, in direct analogy with the homothetic case.

However, we shall not proceed with this type of analysis here since we
wish to determine and display all the kinematical self-similarities in conventional forms; i.e., in the form corresponding to (4.48) and in a form that explicitly displays the self-similar form of the solutions.

In order to derive the spacetimes in manifestly self-similar 
form, we write the metric in comoving spherically symmetric coordinates
adapted to $\pmb{\xi}$, viz.,
$$ds^2 = -e^{2 \phi (\xi)} dt^2 + e^{2 \psi(\xi)} dr^2 + r^2 S^2 (\xi) d \Omega^2, \tag4.50$$
where $d \Omega^2 \equiv d \theta^2_1 + \sin^2 \theta_1 d \theta^2_2$, and
$${\bold u} = e^{- \phi} \frac{\pmb{\partial}}{\pmb{\partial} t}, \tag4.51 $$
where, assuming $\pmb{\xi}$ is neither parallel to nor orthogonal to $\bold u$,
the kinematic self-similar vector $\pmb{\xi}$ and the corresponding self-similar
variable $\xi$ can be written in one of the following forms (Carter and 
Henriksen, 1989):
$$\align
\text{first kind } (\alpha =1; HV): \qquad  & \pmb{\xi} = t \frac{\pmb{\partial}}{\pmb{\partial} t} + r \frac{\pmb{\partial}}{\pmb{\partial} r}; \qquad \xi = r/t, \tag 4.52\\
\text{second kind } (\alpha \neq 0, 1): \qquad  & \pmb{\xi} = \alpha t \frac{\pmb{\partial}}{\pmb{\partial} t} + r \frac{\pmb{\partial}}{\pmb{\partial} r}; \quad \xi = r/(\alpha t)^{1/\alpha}, \tag 4.53\\
\text{zeroth kind } (\alpha = 0): \qquad & \pmb{\xi} = \frac{\pmb{\partial}}{\pmb{\partial} t} + r \frac{\pmb{\partial}}{\pmb{\partial} r}; \qquad \xi = re^{-t}. \tag4.54
\endalign $$

Now, if the acceleration is zero, we have that $e^\phi = a_0$, a constant, and if the shear is zero, we have that 
$e^\psi = f_o S$, where
$f_o$ is a constant.  Therefore, the FRW metric can be written in the form
$$ds^2 = - a^2_o dt^2 + S^2 (\xi) [f_o^2 dr^2 + r^2 d \Omega^2]. \tag4.55 $$

\noindent
(i) If $\alpha \neq 0$, and assuming $S' \neq 0$ (non-static case), the EFEs (for a perfect fluid source) then yield
$$S=d\xi^c, \tag4.56 $$
and $(c \neq -1)$
$$f^2_o = (1+c)^2. \tag4.57 $$

The forms for $\pmb{\xi}$ in (4.52) and (4.53) are invariant under the
changes $t \rightarrow at$ and $r \rightarrow br$, and $a$ and $b$ 
can be chosen
to set $a_o=1$ and $d =1$, and we can then write the metric as
$$ds^2 = -dt^2 + \xi^{2c} \left\{\frac{dr^2}{(1+c)^2} + r^2 d \Omega^2  \right\}.\tag4.58 $$
Recalling that $\xi = r (\alpha t)^{-1/\alpha} (\alpha \neq 0)$, and defining a new radial 
coordinate $\bar{r}$ by $\bar{r} = \alpha^a r^{c+1}$, where $a \equiv -c/\alpha$, the FRW metric takes on
its familiar form
$$ds^2 = -dt^2 + t^{2a}  \{ d\bar{r}^2 + \bar{r}^2 d \Omega^2\}, \tag4.59 $$
where $\pmb{\xi}$ is now given by
$$\pmb{\xi} = \alpha t \frac{\partial}{\partial t} + (1+c) \bar{r} \frac{\partial}{\partial \bar{r}}. \tag4.60 $$

We note that each orthogonal $3$-space is flat
$(k=0)$ and that the simple power-law solutions give rise 
to an 
equation of state of the form $p = (\gamma -1) \mu$, and equations (1.12) are automatically satisfied.  The known homothetic 
case $(\pmb{\xi} = \bold H$ for $k =0$) is included here.  In fact, each spacetime (4.59) 
(for each value of $a$) admits kinematical self-similar vectors
for all values of $\alpha \, (\neq 0)$.  [This is known to be true in 
the dust subcase (Carter and Henriksen, 1989)].  This, of course, arises due to the separability of $S(\xi)$ in 
(4.55) (e.g., see equations (4.53) and (4.56)).

\noindent
(ii)  If $\alpha =0$, then the EFEs imply that
$$S = c \xi, \tag4.61 $$
and 
$$f_o \, ^2 = 4. \tag4.62$$
Since $\xi = re^{-t}$, under the  transformation $t \rightarrow t +b$, $b$ can be chosen so that $c =1$, and hence
$$ds^2 = -a^2_o dt^2 + r^2 e^{-2t} (4dr^2 + r^2 d \Omega^2). \tag4.63 $$
Finally, defining the new variables $\bar{t} = a_ot, \bar{r} = r^2$,
the metric becomes
$$ds^2 = -d\bar{t}^2 + e^{-2\bar{t}/a_o} (d\bar{r}^2 + \bar{r}^2 d \Omega^2),   \tag4.64 $$
and the kinematical self-similar vector is given by
$$\pmb{\xi} = a_o \frac{\pmb{\partial}}{\pmb{\partial} \bar{t}} + 2 \bar{r} \frac{\pmb{\partial}}{\pmb{\partial}\bar{r}}.   \tag4.65 $$
Again, this FRW model is flat $(k =0)$. However, this case has
no homothetic vector analogue.  The equation of state for this FRW model is given by
$\mu = -p = 12a_o^{-2}$, and the metric is, of course, the 
de Sitter metric.  Hence, the
de Sitter model admits a self-similarity of the zeroth kind.  Equations (1.12) are trivially satisfied with $a =b =0$.

Finally, we must deal with the two special cases not 
considered above in which $\pmb{\xi}$ is either parallel to or orthogonal to $\pmb{u}$.  For illustrative purposes let us return to the more conventional coordinate system in which the metric is given by (4.48).  Writing $\pmb{\xi}$
in the form
$$\pmb{\xi} = \pmb{\xi}^o (r,t) \frac{\pmb{\partial}}{\pmb{\partial} t} + \xi^r (r,t) \frac{\pmb{\partial}}{\pmb{\partial }r}, \tag4.66 $$
equations (1.10) and (1.11) yield
$$\align 
\xi^o & = \alpha t + \beta, \tag4.67\\
\xi^r & = \frac{(2 -c)}{2} r, \tag4.68 
\endalign$$
where $c$ is an arbitrary constant, and
$$\frac{2 \dot{R}}{R} (\alpha t + \beta) = c, \tag4.69$$
and
$$k(c-2) =0. \tag4.70 $$
{}From this last equation we see that either $k =0$, and the 
FRW spacetime is flat, or $c=2$, whence the kinematic self-similar 
vector is parallel to the fluid velocity vector (i.e., $\xi^r =0$).  Let us 
consider this latter case first.

\noindent
(a)  $\pmb{\xi}$ parallel to $\bold u$.  In this case 
$c=2$ (i.e., $\xi^r =0$) and $k$ is unrestricted 
by equations (4.67)--(4.70), whence equation (4.69) becomes 
($\alpha^2 + \beta^2 \neq 0$)
$$\frac{\dot{R}}{R}(\alpha t + \beta) =1. \tag4.71$$

If $\alpha \neq 0$, we can set $\beta =0$ by a time translation, whence 
equation (4.71) yields
$$R = dt^{1/\alpha}, \tag4.72 $$
so that when $\alpha =1$ (homothetic vector case) we recover the usual
FRW model with $\mu + 3p = 0$ (valid for all $k$), whereas for
$\alpha \neq 1$ all power law solutions of the form (4.72) admit
a vector field satisfying equations (1.10) and (1.11).

We note that if $k =0$, then the spacetime (4.59) also admits 
additional kinematic self-similarities parallel to $\bold u$ 
(in addition to those given by (4.60)).  If, on the other hand,
$k \neq 0$, we can see that the resulting FRW models with power law 
scale factors always admit a vector field
parallel to the fluid  velocity vector that satisfies equations
(1.10) and (1.11).  However, equations (1.12) need not be satisfied.  Indeed,
from equation (4.72) we find that
$$\align
\mu & = \frac{3}{\alpha^2} t^{-2} + \frac{3k}{d^2} t^{-2/\alpha},  \tag 4.73\\
p & = \frac{1}{\alpha}(2 - 3/\alpha) t^{-2} - \frac{k}{d^2}t^{-2/\alpha},\tag4.74
\endalign $$
so that equations (1.12) can only be satisfied if (either)
$\alpha =1$, whence $\pmb{\xi}$ is a homothetic vector and $\mu + 3p =0$ as 
before (or $k =0$).

It is curious to note, however, that if we consider
the perfect fluid source to be due to two separate comoving perfect fluids (Coley and Tupper, 1986), so that
$$\aligned
 \mu & = \mu_1 + \mu_2, \\
p & = p_1 + p_2, 
\endaligned \tag4.75 $$
where
$$\mu_1 = \frac{3}{\alpha^2} t^{-2}; \quad p_1 = (2 \alpha/3 -1) \mu_1 \tag4.76$$
and
$$\mu_2 = \frac{3 \kappa}{d^2}t^{-2/\alpha}; \quad p_2 = -\frac{1}{3} \mu_2  \tag4.77 $$
then
$$\align
{\Cal L}_{\pmb{\xi}_{||}} \mu_1 & = -2 \alpha \mu_1, \quad {\Cal L}_{\pmb{\xi}_{||}} p_1 = -2 \alpha p_1\\
{\Cal L}_{\pmb{\xi}_{||}} \mu_2 & = -2 \mu_2, \qquad {\Cal L}_{\pmb{\xi}_{||}} p_2 = -2p_2 \tag4.78
\endalign $$
where ${\pmb{\xi}}_{||} = \alpha t \frac{\pmb{\partial}}{\pmb{\partial}t}$, so that each separate fluid satisfies
equations of the form (1.12).

If $\alpha = 0$, then equation (4.71) yields
$$R = de^{t/\beta}  .\tag4.79$$
That is, FRW spacetimes with (4.79) admit a vector field
$\pmb{\xi}_{||} = \alpha t \frac{\pmb{\partial}}{\pmb{\partial} t} = \alpha t \bold u$ satisfying equations (1.10) and (1.11).  If $k =0$, 
then spacetimes of the form (4.64) admit additional kinematic
self-similarities parallel to $\bold u$.  There is no analogous result in the
case of a homothetic vector.  If $k \neq 0$, however, then equations (1.12) cannot
be satisfied.  Again, if we consider the two separate comoving
perfect fluids interpretation (4.75) with
$$\align
\mu_1 & = \frac{3}{\beta^2} = -p_1, \tag4.80 \\
\mu_2 & = \frac{3k}{d^2} e^{-2t/\beta} = -3p_2, \tag4.81
\endalign $$
then each separate fluid satisfies equations of the form (1.12).

If $c \neq 2$ ($\xi^r \neq 0$), then from equation (4.70) we must have $k =0$; i.e., the
FRW models are of zero curvature.  If $\alpha^2 + \beta^2 \neq 0$, then solving
equation (4.69) yields the spacetimes 
(4.59) and 
(4.64) obtained earlier.  If $\alpha = \beta = 0$, then $\xi^o = 0$.

\noindent
(b)  $\pmb{\xi}$ orthogonal to $\bold u$.  If $\alpha = \beta =0$, then $\xi^o =0$ and
the vector $\pmb{\xi}$ is orthogonal to the 
fluid  velocity vector.  Equation (4.69) simply 
yields $c =0$, so that
$$\pmb{\xi}_\perp = r \frac{\pmb{\partial}}{\pmb{\partial} r}; \tag 4.82 $$
that is, every flat FRW model admits a
vector field of the form (4.82) which satisfies the
conditions (1.10) and (1.11); in fact,
$${\Cal L}_{\pmb{\xi}_\perp} u_a =0.  \tag4.83 $$
In addition, since $\mu = \mu(t)$ and $p = p(t)$, it follows
immediately that
$${\Cal L}_{\pmb{\xi}_\perp} \mu =0 = {\Cal L}_{\pmb{\xi}_\perp} p,  \tag4.84 $$
so that equations (1.12) are trivially satisfied.  Hence, every
\underbar{flat} FRW model (for any $R(t))$ 
admits a vector field $\pmb{\xi}_\perp$, given by (4.82), orthogonal to $\bold u$, that satisfies equations (1.10), (1.11), (1.12); in particular,
${\Cal L}_{\pmb{\xi}_\perp} h_{ab} = 2h_{ab}$ and 
equations (4.83) and (4.84) are satisfied.

\head 5. Conclusion \endhead

After a brief review of self-similarity and its 
applications in general relativity, the covariant notion
of a kinematic self-similarity in the context of relativistic
fluid mechanics was introduced (Carter and Henriksen, 1989), which it was 
argued is the
natural relativistic counterpart of self-similarity of the more
general second (or zeroth) kind and hence is a generalization
of a homothety which corresponds to self-similarity of the
first kind (Cahill and Taub, 1971).  Various mathematical and
physical properties of spacetimes admitting a kinematic
self-similarity were discussed.  We note the relationship between
a kinematic self-similarity and what Collins and Szafron (1979) term
an intrinsic symmetry and what Tomita (1981) refers to as a partial homothety (however, see also Tomita and Jantzen, 1983).

The governing equations (adopted from Ellis, 1971)
of the perfect fluid (cosmological) models under investigation were 
introduced, and a set of integrability conditions for the
existence of a proper kinematic self-similarity in the spacetime models 
was derived.     These important constraints,
the integrability conditions, given by equations (2.16) in general
and by equations (2.27) - (2.30) in the particular case of
a perfect fluid, played a central role in the resulting
analysis.  All of the relevant equations were then given in
the physically important case of zero vorticity.

\noindent
{\bf A. Summary}

Exact solutions of the irrotational perfect fluid
Einstein field equations admitting a kinematic
self-similarity were then sought in a number of special cases.
Since the integrability conditions constitute very severe constraints, such 
solutions are necessarily
of a particularly simple form.

First the geodesic  (i.e., zero acceleration) case 
was considered in subsection 3.A.  It was proven that (provided the shear is
non-zero)  the $3$-spaces orthogonal to ${\bold u}$ are Ricci-flat,
that is,  $^3\!R_{ab} =0.$
This case consequently merits further consideration since there
have been various studies of spacetimes with vanishing
$3$-Ricci tensor (see, for example, Collins and Szafron, 1979, and
Stephani and Wolf, 1986). It was further proven that $a = -2 \alpha$, 
and the form of the kinematic self-similar vector $\pmb{\xi}$ and the
pressure were given in the particular coordinates (3.1)/(3.2) by equations
(3.16) and (3.18), respectively $(a = b =0$ and $p =$ constant in the
special case $\alpha =0$).  The expansion was shown to be
governed by the differential equation (3.23)/(3.24), which was
in fact integrated in the special case $\alpha =0$ [see equation
(3.36)].

The further specialization to dust (i.e., zero pressure)
was then considered in subsection 3.B. In this case the governing
differential equation (3.23) reduces to equation (3.32)
which can be completely integrated to obtain the
expansion (see eqns. (3.40) - (3.43)).  The asymptotic
properties of this class of solutions was studied, and it
was found that the resulting models are asymptotic (at
late times) to an exact flat FRW (Einstein-de Sitter) model and are asymptotic
(at early times) to an exact vacuum ($3$-Ricci flat) model
with $3\sigma^2 =\theta^2 = t^{-2} -$  for a large class of models this was
shown to be the Kasner (Bianchi I) solution.
We note that these exact (asymptotic) solutions are known 
to admit a {\it homothetic vector}.

The case of zero expansion, studied in subsection 3.C,
was shown to subdivide into two subcases.  In the first subcase
the models are necessarily shear-free with zero magnetic part
of the Weyl tensor, and in general $p = (\gamma-1)\mu$ (and
equation (3.94) is satisfied) and the resulting models are
completely static.  In the second subcase the models are
necessarily stiff $(p =\mu)$ non-static perfect
fluid models.  In the coordinates (3.61)/(3.62) the form of 
the kinematic self-similarity was found to be given by equation (3.65)
and either equation (3.95) or (3.106).

In subsection 4.A perfect fluid spacetimes admitting a
kinematic self-similar vector $\pmb{\xi}$ parallel to the velocity
vector $\bold u$ were studied, and it was proven that such
spacetimes are necessarily FRW spacetimes with $p = [\frac{2 \alpha}{3} -1]\mu$
and $\mu = \frac{1}{3} \theta^2 = \frac{3}{\alpha^2} t^{-2}$.  In the case that 
$\pmb{\xi}$ is orthogonal to
${\bold u}$, it was shown in subsection 4.B
that in general $[\alpha b(\alpha +b) \neq 0] \enskip p = (\gamma-1) \mu$ with $\gamma =b/(\alpha +b)$
and $\theta$ satisfies the differential equation (4.40)
(and $a =b)$, and that either $a = -2$ whence equations (4.42) - (4.44)
follow or $a = -2 \alpha$ and equations (4.45) - (4.47) result.

Finally, in subsection 4.C, the existence of
kinematic self-similarities were studied in FRW spacetimes.
In the general case in which $\pmb{\xi}$ is neither parallel nor orthogonal
to $\bold u$, it was shown that if $\alpha \neq 0$ then the resulting FRW model is flat
with $p = (\gamma -1)\mu$ and the scale factor is of a simple power-law form
[and $\pmb{\xi}$ is given by (4.60) in conventional coordinates].  Indeed, each such
FRW model admits kinematic self-similar vectors for all 
values of $\alpha \, (\neq 0)$ in addition to a homothetic vector.  On the other
hand, if $\alpha =0$ the FRW model is necessarily a flat de Sitter
model with $\mu + p =0$ [and $\pmb{\xi}$ is given by (4.65)].  Hence
the de Sitter spacetime admits a self-similarity of the zeroth
kind (note that such a spacetime cannot admit a homothety).
In addition, all FRW models with a scale factor of the 
power-law form (4.72) with $\alpha \neq 0$ [and with a scale factor of the exponential
form (4.79) when $\alpha =0$] were shown to admit a vector field $\pmb{\xi}$ parallel to $\bold u$
which satisfies equations (1.10) and (1.11).  Here the curvature need 
not be zero; however, in the flat case the FRW models
admit these special vector fields (parallel to $\bold u$) in addition to the kinematic
self-similar vectors mentioned above.  Finally, every flat
FRW model (with a  scale factor of any form) was shown to admit a 
vector field $\pmb{\xi}$ of the form (4.82), orthogonal to $\bold u$,
which satisfies equation (1.10) and equations (1.11) and (1.12) [see
equations (4.83) and (4.84)].

\noindent
{\bf B. Discussion}

There are a variety of circumstances in general relativity theory, and particularly in cosmology, 
in which self-similar models act as asymptotic states of more 
general models.  Indeed, in a number of classes of perfect fluid cosmological
models with equation of state $p = (\gamma -1) \mu$ and in which the 
governing equations reduce to a dynamical system, including,
for example, spatially homogeneous models and silent universe
models, and in some cases spherically symmetric models and
$G_2$ models, it is known  
that exact solutions admitting a homothetic vector
play an 
important role in describing the asymptotic properties of these
models (see Coley, 1996, for a review and appropriate references).

For example, orthogonal spatially homogeneous models
have attracted much attention since the governing equations
reduce to a relativity simple finite dimensional system of
autonomous ordinary differential equations (Wainwright and Ellis, 1996 - 
henceforward WE).  Wainwright and collaborators 
(see Refs. in WE and Coley, 1996) have utilized an orthonormal frame approach
and introduced an expansion-normalized (and hence dimensionless) set of
variables to study these models.  In particular, it was proven that
all the singular points of the (corresponding ``reduced'') system
(of ordinary differential equations) correspond to exact solutions
admitting a homothetic vector (Hsu and Wainwright, 1986).
It is in this sense that self-similar models play an
important role in describing the dynamics of
spatially homogeneous models
asymptotically. [The situation is complicated by the fact that in the
more general classes of Bianchi models there exist
more complicated attractors than simple singular points; for
example, in models of Bianchi type IX (and VIII) there is oscillatory
behaviour with chaotic-like characteristics as one
follows the evolution into the past towards the initial singularity
due to the existence of a $2$-dimensional attractor in
the $5$-dimensional phase space (WE).]

In addition, in the class of inhomogeneous $G_2$ 
cosmological models (in which the spacetime admits two commuting
spacelike Killing vectors acting orthogonally transitively) it has been
shown by Hewitt and Wainwright (1990) that the governing
Einstein field equations can be written as an
(infinite dimensional) autonomous system of first-order quasi-linear partial differential
equations in terms of two independent dimensionless variables, and it was proven that
the associated dynamical equilibrium states correspond to exact cosmological
solutions that admit a homothetic vector (and which are consequently self-similar).
In a particular subclass of $G_2$ cosmologies, the separable diagonal $G_2$ models,
it was shown that the models do indeed asymptote towards the 
dynamical  equilibrium points, and Wainwright and Hewitt
have conjectured that this may be the case for  more general $G_2$
models.  Hence self-similar models may play an important
role in describing the asymptotic dynamical
behaviour of these inhomogeneous cosmological models.

In this paper we have studied models which admit a kinematic self-similar
vector.  We note that in all cases in which we have either been able
to integrate the equations to obtain exact solutions or we
have been able to determine the asymptotic behaviour of a
class of models, the asymptotic behaviour has been represented by
an exact solution admitting a proper homothetic vector (and hence a
model which is self-similar of the first kind).  For example, in the pressure-free
case studied in subsection 3.B, models were found to be asymptotic to the
Einstein-de Sitter model (to the future) and the Kasner model (to the past), and
both of these
exact solutions admit a homothetic vector.  In addition, the 
particular, exact, perfect fluid spherically symmetric solutions studied by Benoit and
Coley (1996) [which include the dust solutions of 
Lynden-Bell and Lemos (1988) and Carter and Henriksen (1989)]
were shown to be asymptotic (both to the past and to the future)
to exact FRW models which admit a homothetic vector, and it was also
shown in Benoit and Coley (1996) that the equilibrium
points at finite values of the autonomous system of ordinary
differential equations which govern the general class of perfect fluid
spherically symmetric, kinematic self-similar 
models correspond to exact solutions which admit a homothetic vector.
It appears that the same is true for the analogous solutions
in the case of plane symmetry  (work in progress).

It would be interesting to determine whether this is true in
more generality.  That is, {\it it is an interesting and important question}
{\it to determine the conditions under which models admitting a proper}
{\it kinematic self-similarity are asymptotic to an exact homothetic
solution}.  This might then shed light on when self-similar models
of the second kind play an important role in determining the
``intermediate asymptotic'' behaviour of solutions (Barenblatt and
Zeldovich, 1972), and the role of generalized self-similar models
in describing the asymptotic properties of models.

The exception to the above is the de Sitter model.  This flat FRW 
model with equation of state $\mu =-p=\text{ constant }$  (or equivalently with a
cosmological constant) does not admit a homothetic vector and is not
asymptotic to a solution that does. Hence it cannot be true that
kinematic self-similar models are asymptotic to exact homothetic
solutions under all circumstances.

The de Sitter model does, however, admit a self-similarity of
the zeroth kind (see subsection 4.C).  It is curious to note that the
de Sitter model acts as an asymptotic state of more general
models (according to the so-called cosmic no-hair theorems).  In particular, it was shown
by Wald (1983) that spatially homogeneous Bianchi models with
a cosmological constant (except those of Bianchi type IX which
recollapse) are future asymptotic to the de Sitter model.  The 
existence of the de Sitter model as a future asymptotic state
is indicative of exponential inflation.  Hence, {\it it is of}
{\it interest to determine under what conditions the asymptotic states}
{\it of cosmological models are represented by solutions of Einstein's
field equations admitting a generalized self-similarity} (i.e., not
just a homothety).  This question deserves to be studied further.

Recently Wainwright's work has been generalized to the case of 
imperfect fluid Bianchi models
satisfying the non-causal linear Eckart theory
of irreversible thermodynamics (Coley and van den Hoogen, 1994)
and the causal theory (both the truncated version and the full theory)
of Israel and Stewart (see Coley et al., 1996, and references within).
Dimensionless physical variables (similar to those used by Wainwright)
were utilized and a set of ``dimensionless equations of state'' were assumed,
whence it was again shown that in general the singular points of the resulting
dynamical system are represented by exact homothetic solutions (Coley and van den Hoogen, 1994).  In the
exceptional cases the singular points correspond to models which
violate the strong energy conditions and have constant expansion, and the
models are analogues of the de Sitter solution (with the viscous
terms mimicking a cosmological constant) and are consequently
self-similar of the zeroth kind.  

Moreover, in other work viscous
fluid models have been studied (particularly in the case of simple FRW and
Bianchi spacetimes) in which the governing equations reduce to a
(simple) system of autonomous ordinary differential equations,
but since particular equations of state were assumed
that are not of a ``dimensionless'' form the associated
singular points do not necessarily correspond to  exact solutions
admitting a homothetic vector. [In this work the viscosity
coefficients are modeled by both the non-causal Eckart
and causal Israel-Stewart theories of irreversible thermodynamics;
the reader is directed to the research papers of the Polish and
Russian groups and the Russian and Spanish groups, respectively,
which are fully referenced in the papers by Coley and collaborators cited here.]
In a preliminary investigation of this work it appears that all singular points
correspond either to an exact solution which is known to admit a homothetic
vector, to an exact zero-curvature FRW model (not
necessarily admitting a homothety), or to a de Sitter-like solution.
Hence, all of these models appear to be asymptotic to 
models which admit a kinematic self-similarity of the
zeroth, first or second kind.  Clearly this needs to be
studied further.

However, {\it it is clear that kinematic self-similar models play an 
important role in describing the asymptotic properties of cosmological models.}
It is interesting to ask
to what extent are cosmological models (or, rather, what is
the class of solutions of Einstein's field equations which are)
asymptotic to self-similar solutions, when self-similarity
is understood in its more general sense.

\heading{Appendix: The vacuum case.}\endheading

Vacuum spacetimes admitting a homothetic vector were studied
by McIntosh (1975), in which it was shown that a non-flat vacuum
spacetime can only admit a non-trivial homothetic vector if that homothety 
is non-null and is 
not hypersurface orthogonal.

The study of kinematic self-similar vectors in vacuum spacetimes
is not physically well-motivated, since in general there does not exist
a physically or intrinsically defined timelike vector $\bold u$ with respect to which the metric can be 
uniquely decomposed (unlike in the case of a perfect fluid spacetime, for example, in
which there exists such a vector which has both an intrinsic physical and geometrical role \-- it is both tangent
to the fluid flow and is the unique normalized timelike eigenvector of the Ricci tensor), 
and consequently there exists no such $\bold u$ with respect to which the definition (1.10)/(1.11) can be applied.  However, for curiosities sake, let us 
study the consequences of the existence of an intrinsically
defined timelike vector $\bold u$ which satisfies equations (1.10) and (1.11) 
in a vacuum spacetime.

{}From subsection 2B (for $\omega =0$), in the case  $\mu = p=0 $ we obtain 
the equations
$$\dot{u}^c\, _{;c} = 0, \tag A.1 $$
and 
$$\dot{\theta} + \theta^2 = 0, \tag A.2$$
and hence 
$$\sigma^2 = \frac{1}{3} \theta ^2, \tag A.3$$
and consequently 
$$^3\!R =0. \tag A.4$$

Let us adopt coordinates so that the metric functions are
defined through (3.61) and equations (3.62) and (3.63) are valid, whence on
defining the new time parameter, $\tau$, by
$$\Phi' \equiv \frac{\partial \Phi}{\partial \tau} \equiv \frac{1}{\theta 
U} \, \frac{\partial \Phi}{\partial t},  \tag A.5 $$
we can then integrate equation (A.2) to obtain
$$\theta = \Theta (x^\gamma)e^{-\tau}, \tag A.6 $$
whence equation (A.3) yields
$$\sigma^2 = \frac{1}{3} \Theta^2 e^{- 2 \tau}, \tag A.7$$
which also follows from (2.37) [when $\sigma \neq 0$; if $\sigma =0$, then if follows that $\theta =0$].
By definition we have that
$$H^{\alpha \beta} H^\prime_{\alpha \beta}   = 2,   \tag A.8 $$
and
$$\sigma_{\alpha \beta} = \frac{\theta}{2} H^\prime_{\alpha \beta}  - \frac{\theta}{3} H_{\alpha \beta},  \tag A.9 $$
and
$$^3\!R_{\alpha \beta} = \frac{1}{U} U_{, \alpha\beta} - \frac{1}{U} \Gamma^\gamma_{\alpha \beta} \, U_{, \gamma} \tag A.10 $$
(all other components of $\sigma_{ab}$ and $^3\!R_{ab}$ vanish).  Finally, 
equation 
(2.37) yields
$$\sigma'_{\alpha \beta} - H^{\gamma \delta} \sigma_{\gamma(\alpha}\!H^\prime_{\beta)\sigma}  + \sigma_{\alpha \beta} = 0,     \tag A.11$$
and using (A.9) we consequently obtain
$$H^{\prime \prime}_{\alpha \beta} - H^{\gamma \delta} H^\prime_{\alpha \gamma}  H^\prime_{\beta \delta}  = 0. \tag A.12$$

In the case that $H_{\alpha \beta}$ is diagonal, i.e.,
$H_{\alpha \beta}(\tau,x^\gamma) \equiv \text{diag} \{h_1, h_2, h_3\}$, we can integrate
equations (A.12) to obtain
$$h_\nu(\tau, x^\gamma) = G_\nu(x^\gamma) e^{2 F_\nu(x^\gamma)\tau} \enskip (\nu = 1, 2, 3), \tag A.13$$
whence equation (A.8) then implies that
$$F_1 + F_2 + F_3 = 1. \tag A.14$$
We note that the spatially homogeneous vacuum 
Bianchi I (Kasner) model is a particular solution of equations
(A.2) - (A.4), (A.8) and (A.12) in which $U=1$ (and hence
$^3\!R_{ab} =0)$.

\heading{Acknowledgments}  \endheading

I would like to thank Patricia Benoit for reading the manuscript and I would like  to acknowledge
the Natural Sciences and Engineering Research Council of Canada for 
financial support.
 
\heading{REFERENCES}\endheading

\roster

\item"" G. I. Barenblatt, 1952, Prikl. Mat. Mekh. {\bf 16}, 67.

\item"" G. I. Barenblatt and Ya B. Zeldovich, 1972, Ann. Rev. Fluid 
Mech. {\bf 4}, 285.

\item"" P. M. Benoit and A. A. Coley, 1996, Class. Q. Grav., submitted.

\item""  E. Bertschinger, 1985, Ap. J. {\bf 268}, 17.

\item"" A. H. Cahill and M. E. Taub, 1971, Comm. Math. Phys. {\bf 21}, 1.

\item"" B. Carter and R. N. Henriksen, 1989, Ann. Physique Supp. {\bf 14}, 47.

\item""  B. Carter and R. N. Henriksen, 1991, J. Math. Phys. {\bf 32}, 2580.

\item"" J. Castejon-Amenedo and A. A. Coley, 1992, Class. Q. Grav. {\bf 9}, 2203.

\item"" A. A. Coley, 1991, Class Q. Grav. {\bf 8}, 955.

\item"" A. A. Coley, 1996, in the {\it Proceedings of the Sixth Canadian Conference on 
General Relativity and Relativistic Astrophysics}, the Fields Institute Communications Series (AMS), eds. S. P. Braham, J. D. Gegenberg and 
R. J. McKellar (Providence, RI). 

\item"" A. A. Coley and R J. van den Hoogen, 1994, J. Math Phys. {\bf 35}, 4117.

\item"" A. A. Coley, R J. van den Hoogen and R. Maartens, 1996, Phys. Rev. D. {\bf 54}, 1393.

\item"" A. A. Coley and D. J. McManus, 1994, Class. Q. Grav. {\bf 11}, 1261.

\item"" A. A. Coley and B. O. J. Tupper, 1986, J. Math. Phys. {\bf 27}, 406.

\item"" C. B. Collins and D. A. Szafron, 1979, J. Math. Phys. {\bf 20}, 2347.

\item"" L. Defrise-Carter, 1975, Comm. Math. Phys. {\bf 40}, 273.

\item"" D. M. Eardley, 1974, Comm. Math. Phys. {\bf 37}, 287.

\item"" G. F. R. Ellis, 1971, {\it Relativistic Cosmology}, in 
{\it General Relativity and Cosmology, XLVII Corso, Varenna, Italy} (1969), ed. R. Sachs (Academic, New York).

\item"" S. W. Goode and J. Wainwright, 1982, Phys. Rev. {\bf D 26}, 3315.

\item"" R. N. Henriksen, 1989, MNRAS {\bf 240}, 917.

\item"" R. N. Henriksen, A. G. Emslie and P. S. Wesson, 1983, Phys. Rev. D {\bf 27}, 1219.

\item"" C. G. Hewitt and J. Wainwright, 1990, Class. Quantum Grav. {\bf 7}, 2295.

\item"" L. Hsu and J. Wainwright, 1986, 
Class. Quantum Grav. {\bf 3}, 1105.

\item"" S. Ikeuchi, K. Tomisaka and J. P. Ostriker, 1983, Ap. J. {\bf 265}, 583.

\item"" D. Kramer, H. Stephani, M. A. H. MacCallum and E. Herlt, 
1980, {\it Exact Solutions of Einstein's Field Equations}
(Cambridge University Press, Cambridge).

\item""  D. Lynden-Bell and J. P. S. Lemos, 1988, MNRAS {\bf 233}, 197.

\item"" R. Maartens and S. D. Maharaj, 1986, Class, Q. Grav. {\bf 3}, 1005.

\item"" C. B. G. McIntosh, 1975, Gen. Rel. Grav. {\bf 7}, 199.

\item"" J. Ponce de Leon, 1993, Gen. Rel. Grav. {\bf 25}, 865.

\item"" J. Schwartz, J. P. Ostriker and A. Yahil, 1975, Ap. J. {\bf 202}, 1.

\item"" L. I. Sedov, 1967, {\it Similarity and Dimensional Methods in Mechanics} (New York, Academic).

\item"" H. Stephani and Th. Wolf, 1986, in {\it Galaxies, Axisymmetric Systems and Relativity}, ed. M. A. H. MacCallum, p.275 (Cambridge University Press, Cambridge).

\item"" K. Tomita, 1981, Prog. Theoret. Phys. {\bf 66}, 2025.

\item"" K. Tomita and R. T. Jantzen, 1983, Prog. Theoret. Phys. {\bf 70}, 886.

\item"" J. Wainwright and G. F. R. Ellis, 1996, {\it Dynamical Systems in Cosmology} (Cambridge University Press, Cambridge).

\item"" R. M. Wald, 1983, Phys. Rev. D. {\bf 28}, 2118.

\item"" K. Yano, 1955, {\it The Theory of Lie Derivatives} (North Holland, Amsterdam). 

\endroster 

\enddocument